\documentclass[twocolumn,prb]{revtex4-1}

\usepackage{amsmath,amssymb}
\usepackage{hyperref}

\usepackage{epsfig}
\usepackage{graphicx}
\usepackage{graphics}
\usepackage{mathdots}
\usepackage{url}
\usepackage{epstopdf}

\begin{document}

\title[Vortex ratchet reversal in an asymmetric washboard pinning potential]
{Vortex ratchet reversal in an asymmetric washboard pinning potential subject to combined dc and ac stimuli}

\author{Valerij~A~Shklovskij$^{1,2}$, Vladimir V Sosedkin$^2$, and Oleksandr~V~Dobrovolskiy$^{2,3}$}

\address{$^1$Institute of Theoretical Physics, NSC-KIPT, 61108 Kharkiv, Ukraine\\
        $^2$Physical Department, Kharkiv National University, 61077 Kharkiv, Ukraine\\
        $^3$Physikalisches Institut, Goethe-University, 60438 Frankfurt am Main, Germany}
        
\begin{abstract}
The mixed-state resistive response of a superconductor thin film with an asymmetric washboard pinning potential subject to superimposed dc and ac currents of arbitrary amplitudes and frequency at finite temperature is theoretically investigated. The problem is considered in the \emph{single-vortex} approximation, relying upon the \emph{exact} solution of the Langevin equation in terms of a matrix continued fraction. The dc voltage response and the absorbed power in ac response are analyzed as functions of dc bias, ac current amplitude and frequency, in a wide range of corresponding dimensionless parameters. Predicted are (i)~a reversal of the rectified voltage at small dc biases and strong ac drives and (ii)~a non-monotonic enhancement of the absorbed power in the nonlinear ac response at far sub-depinning frequencies. It is elucidated how and why both these effects appear due to the competition of the fixed \emph{internal} and the tunable, dc bias-induced \emph{external} asymmetry of the potential as the only reason. This is distinct from other scenarios used for explaining the vortex ratchet reversal effect so far.
\end{abstract}

\maketitle

\section{Introduction}
The investigation of directed and reversed net transport in systems lacking reflection symmetry, i.e. in \emph{ratchets}, in the presence of deterministic or stochastic forces with time averages of zero has been a fascinating topic of research over the last two decades~\cite{Rei02prt,Han09rmp,Sil10inb}. The originally mechanical ratchet scenario~\cite{Fey63boo} was successfully adopted in biology while studying molecular motors~\cite{Svo93nat} and has been exploited in many other areas of science and engineering. For instance, directed net transport has been experimentally observed in the motion of colloidal particles~\cite{Rou94nat}, in superconducting quantum interference devices~\cite{Zap96prl}, Josephson junctions~\cite{Car01prl,Knu12pre}, cold atoms~\cite{Gom06prl}, and in the driven motion of domain walls in asymmetrically patterned magnetic films~\cite{Per08prl}. Besides, one often deals with the ratchet effect when investigating stepper motors~\cite{Van11prl} and other mode-locking~\cite{Von09prb} and synchronization~\cite{Zar09pre} phenomena.

Among different types of ratchet systems, the superconducting Abrikosov vortex ratchets~(VRs)~\cite{Plo09tas} have gained especial attention. In essence, the VR is a system where the vortex can acquire a net motion whose direction is determined only by the asymmetry of the periodic pinning potential. The asymmetry of the potential refers to the current direction reversal. Initially, the VR effect has been proposed for driving fluxons out of superconductors~\cite{Lee99nat} and for constructing flux pumps and lenses~\cite{Wam99prl}. Later on, it has been observed by magneto-optical imaging~\cite{Tog05prl,Men07prb}. Most commonly, VRs have been used for modifying the magneto-resistive response of nanostructured superconductors~\cite{Sil10inb}. Accordingly, along with the vortex guiding effect the VR effect represents one of the two most important phenomena~\cite{Sil10inb} stipulating the modern field of research and technology known as \emph{fluxonics}. Similar to the manipulation of electrons in micro- and nano-electronics, fluxonics is based upon the manipulation of flux-line vortices in superconductors, usually by making use of artificially fabricated pinning nanolandscapes.

An even more intriguing phenomenon in the VR dynamics is the switching effect between direct and reversed net motion, also known as the \emph{ratchet reversal} effect. Experimentally, a sign change in the rectified voltage as a function of an ac driving force has been widely reported~\cite{Vil03sci,Sil06nat,Din07prb,Gil07prl,Din07njp,Jin10prb,Lar10prb,Lar11prb}. Also, a reversal of the current of particles has been observed in an optical ratchet~\cite{Arz11prl}. For explaining the VR reversal different mechanisms have been proposed. These include the presence of interstitial vortices~\cite{Ols05pcs,Lar11prb}, a reconfiguration of the vortex lattice at different drive and field values~\cite{Din07njp}, the competition of the characteristic vortex-vortex interaction length scale with the period of an asymmetric pinning landscape~\cite{Gil07prl}, the inertia effect~\cite{Jin10prb}, the interaction between vortices within pinning centers~\cite{Sil06nat}, and the coexistence of pinned and interstitial vortices moving in opposite directions~\cite{Vil03sci}. At the same time, it has been shown that VR reversals can also occur when interstitial vortices are absent~\cite{Din07prb}.

To experimentally investigate the VR dynamics, different approaches have been used for tailoring the pinning in an asymmetric fashion. This is achieved, e.g., by using pinning sites of different sizes~\cite{Sil06nat} and shapes, such as triangles~\cite{Vil03sci}, grading circles~\cite{Gil07prl}, and arrow-shaped wedged cages~\cite{Tog05prl}. However, the resulting pinning potential in such structures is rather complex and does not allow for a full and exact theoretical description of the vortex flow at any arbitrary angle with respect to the guiding direction of the potential, as a function of all driving parameters of the problem. At the same time, this complexity is absent for a more simple type of an asymmetric periodic pinning potential ---  the washboard pinning potential (WPP). It is worth noting that already in 1970 Morrison and Rose~\cite{Mor70prl} used In--Bi foils imprinted with a diffraction grating providing an asymmetric WPP landscape. Though the investigation of the VR effect was beyond the scope of their work~\cite{Mor70prl}, the critical current anisotropy caused by to the guiding of vortices along the imprinted channels was clearly seen.

\begin{figure}
    \centering
    \includegraphics[width=0.4\textwidth]{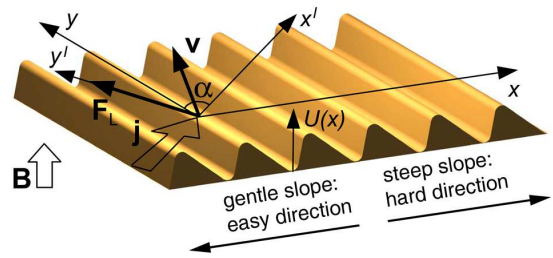}
    \caption[Coordinate system in the rocking-ratchet problem]
    {The coordinate system $xy$ is associated with the WPP channels which are parallel to $\mathbf{y}$ indicating the guiding direction of the WPP. The coordinate system $x'y'$ is associated with the direction of the transport current density vector $\mathbf{j}=\mathbf{j}^{dc}+\mathbf{j}^{ac}\cos\omega t$ directed at the angle $\alpha$ with respect to the $y$ axis. $\textbf{B}$ is the magnetic field and $\mathbf{F}_L$ is the Lorenz force for a vortex. Owing to the average pinning force $\langle \mathbf{F}_p\rangle$ due to the WPP, the average vortex velocity vector $\mathbf{v}$ is not perpendicular to $\mathbf{j}$. The sketch is drawn neglecting the Hall effect. The ratchet WPP $U_p(x)$ is shown to indicate the orientation of its steep-slope and gentle-slope directions along the $x$ axis. The modification of the WPP in the presence of a dc bias is detailed in the caption of figure~\ref{f04TAtrp}.}
    \label{f04TAsys}
\end{figure}

Among state-of-the-art nanofabrication methods suitable for an accurate realization of the WPP used in this work [see below equation~\eqref{e04TArpp}] two advanced mask-less techniques should be mentioned. First, the direct nano-writing by focused ion beam milling~\cite{Utk08vst,Hut10inp} allows one to fabricate nanogroove arrays~\cite{Dob12njp} inducing a WPP in processed films by vortex length reduction and order parameter suppression. Second, a complementary nanofabrication tool, focused electron beam-induced deposition~\cite{Hut12bjn} of the metal-organic precursor $\mathrm{Co_2(CO)_8}$ can be used for furnishing the surface of superconducting films with an array of ferromagnetic $\mathrm{Co}$ strips~\cite{Dob10sst,Dob11pcs} such that a WPP is provided that influences the vortex motion. Both techniques allow one to fabricate carefully designed nanopatterns in which an asymmetry of the induced WPP is achieved by pre-defining the left and right groove or strip slopes' steepness differently~\cite{Dob12ppa}. As for the terminology used throughout this paper, we will use the notions \emph{steep-slope} and \emph{gentle-slope} WPP directions for referring to the \emph{hard} and the \emph{easy} direction for the vortex motion, respectively. These are sketched in figure~\ref{f04TAsys}.

A considerable amount of theoretical work about the general properties of different types of ratchet systems
exists~\cite{Jun96prl,Mat00prl,Rei02apa,Ols05pcs,Sil06nat}. Depending on the way to bring asymmetry into a system, one can distinguish between the fixed \emph{intrinsic} asymmetry, i.e., that caused by the spatial asymmetry of the potential, and the tunable \emph{extrinsic} asymmetry, usually invoked by an external dc bias. A VR with internal pinning asymmetry is called a \emph{rocking ratchet}. In an earlier work~\cite{Shk09prb}, we have studied this case in the limit of non-interacting vortices, i.e., in the \emph{single-vortex approximation} for an asymmetric saw-tooth WPP. A VR with external pinning asymmetry is known as a tilted-potential or \emph{tilting ratchet}. Previously, also this case has been considered~\cite{Shk11prb} for a cosine WPP in the presence of a dc current. In~\cite{Shk11prb}, the exact expressions for the experimentally accessible values being the dc ratchet voltage and the absorbed ac power have been derived by using the \emph{matrix continued fractions} technique~\cite{Shk08prb}. In the present work our objective is to substantially generalize these two studies~\cite{Shk11prb,Shk09prb} and to theoretically investigate a rocking VR subject to superimposed (dc+ac) stimuli. The analytical treatment of the problem is performed with the help of the matrix continued fraction method which will be extended for an asymmetric WPP, although all expressions become more complicated than in the case of a cosine potential.

The motivation of our study is the following: Firstly, we would like to answer the question whether some new effects arise due to the interplay of the tilt-induced and the intrinsic asymmetry of the WPP. Secondly, we would like to propose an exact analytical description of these effects which have not been theoretically addressed so far. These effects are (i)~the~dc ratchet voltage reversal and (ii)~a~non-monotonic enhancement of the power absorption as a function of dc bias at low frequencies. As will be elucidated below, all these predicted effects appear due to the competition of the aforementioned asymmetries as the only reason. This is distinct from other scenarios~\cite{Ols05pcs,Lar11prb, Din07njp,Gil07prl,Jin10prb,Sil06nat} used for explaining the VR reversal effect so far.

The paper is organized as follows. Section~\ref{c04TAfop} presents the general formulation of the problem and its solution in terms of a matrix continued fraction. A graphical analysis of the calculated dependences follows in section~\ref{c04TAgra}. The limiting case of low frequencies is considered in section~\ref{c04TAalc}, with the focus on the scenario of the VR reversal effect. A general discussion of the obtained results concludes our presentation in section~\ref{c04TAcon}.

\section{Main results}
\label{c04TAfop}
\subsection{Formulation of the problem}
The geometry of the problem is schematically shown in figure~\ref{f04TAsys}. Our theoretical treatment of this system relies upon the Langevin equation for a vortex moving with velocity $\mathbf{v}$ in a magnetic field $\mathbf{B}=\mathbf{n}B$ ($B\equiv|\mathbf{B}|$, $\mathbf{n}=n\mathbf{z}$, $\mathbf{z}$~is the unit vector in the $z$ direction and $n=\pm 1$) having the form~\cite{Shk08prb}
\begin{equation}
        \label{e04TAleq}
        \eta\mathbf{v} + n\alpha_{H}\mathbf{v}\times\mathbf{z} = \mathbf{F}_{L}+\mathbf{F}_{p}+\mathbf{F}_{th},
\end{equation}
where $\mathbf{F}_{L}=n(\Phi_{0}/c)\mathbf{j}\times\mathbf{z}$ is the Lorentz force, $\Phi_{0}$ is the magnetic flux quantum, and $c$ is the speed of light. $\mathbf{j}=\mathbf{j}(t)=\mathbf{j}^{dc}+\mathbf{j}^{ac}\cos\omega t$, where $\mathbf{j}^{dc}$ and $\mathbf{j}^{ac}$ are the dc and ac current density amplitudes and $\omega$ is the angular frequency. $\mathbf{F}_{p}=-\nabla U_p(x)$ is the anisotropic pinning force, where $U_p(x)$ is a ratchet WPP. $\mathbf{F}_{th}$ is the thermal fluctuation force, $\eta$ is the vortex viscosity, and $\alpha_{H}$ is the Hall constant. We assume that the fluctuational force $\mathbf{F}_{th}(t)$ is represented by a Gaussian white noise, whose stochastic properties are defined by the relations $\langle F_{th,i}(t)\rangle=0$, $\langle F_{th,i}(t)F_{th,j}(t') \rangle=2T\eta\delta_{ij}\delta(t-t')$, where $T$ is the temperature in energy units, $\langle...\rangle$ means the statistical average, $F_{th,i}(t)$ with $i=x$ or $i=y$ is the $i$ component of $\mathbf{F}_{th}(t)$, and $\delta_{ij}$ is Kronecker's delta.

The ratchet WPP is modeled by
\begin{equation}
        \label{e04TArpp}
        U_p(x) = (U_p/2) [1-\cos kx + e(1 -\sin 2kx)/2],
\end{equation}
where $k=2\pi/a$. Here $a$ is the period and $U_p$ is the depth of the WPP. In equation~\eqref{e04TArpp} $e$ is the asymmetry parameter allowing for tuning the asymmetry strength.
%A family of potentials calculated by equation~\eqref{e04TArpp} for $e=0.01, 0.5, 1, 2$ is exemplified in the inset of figure~\ref{f04TAeve}.
As is apparent from equation~\eqref{e04TArpp}, $e = 0$ corresponds to a cosine WPP. This limiting case of a symmetric WPP has been studied previously~\cite{Mar90pbc,Cof91prl,Maw99prb,Gol92prb, Shk08prb,Shk11prb}. In the opposite limiting case of $e>1$ a double-well WPP ensues. This can also be accounted for within the proposed approach but will be discussed elsewhere. Here, we will focus on the case $e=0.5$, as representative for the most commonly~\cite{Bar94epl,Han96inc,Zap96prl,Mat00prl,Pop00prl,Zar09pre,Arz11prl} used ratchet potential.

Since the ratchet WPP has no component along the $y$ axis the pinning force $\mathbf{F}_p$ has only an $x$ component
\begin{equation}
    \label{e04TApif}
    \hat{F}_{px} = -\sin\textsl{x} + e\cos2\textsl{x},
\end{equation}
where $\hat{F}_{px}=F_{px}/F_p$, $F_p \equiv U_pk/2$, and $\textsl{x}=kx$ is the dimensionless vortex coordinate.

Then, along the $x$ axis equation~\eqref{e04TAleq} acquires the following form
\begin{equation}
    \label{e04TAprx}
    \hat{\tau}(d\textsl{x}/dt) + \sin\textsl{x} - e\cos\textsl{2x} = \hat{F}_{Lx} + \hat{F}_x,
\end{equation}
where $\hat{\tau}\equiv2\eta D/U_pk^2$ is the relaxation time with $D \equiv 1+\delta^2$, $\delta\equiv n\epsilon$, and $\epsilon \equiv \alpha_H/\eta$ is the dimensionless Hall constant. In equation~\eqref{e04TAprx} $\hat{F}_{Lx} = (F_{Lx}-\delta F_{Ly})/F_p$ is the dimensionless generalized moving force in the $x$ direction. $\hat{F}_{x}=(F_{x}-\delta F_{y})/F_p$ and $\langle \hat{F}_{x}(t)\hat{F}_{x}(t')\rangle=\tau\delta(t-t')$, where $\tau\equiv2\hat{\tau}/g$ and $g=U_p/2T$ is the dimensionless inverse temperature.

From the Langevin equation~\eqref{e04TAleq}, the average vortex velocity components can straightforwardly be written as
\begin{equation}
    \label{e04TAvpx}
    \left\{
    \begin{array}{ll}
    \langle v_y \rangle=F_{Ly}/\eta+\delta\langle v_x\rangle,\\
    \\
    \langle v_x \rangle(t) = \displaystyle\frac{\Phi_{0}j_c}{c\eta D}[j^{dc}+j^{ac}\cos \omega t - \langle \sin \textsl{x}\rangle(t) + e \langle \cos \textsl{2x}\rangle(t)],
    \end{array}
    \right.
\end{equation}
where $j^{dc}\equiv n (j^{dc}_y+\delta j^{dc}_x)/j_c$, $j^{ac}\equiv n (j^{ac}_y+\delta j^{ac}_x)/j_c$, $j^{dc}_y=j^d\cos\alpha$, $j^{dc}_x=j^d\sin\alpha$, $j^d=|\mathbf{j}^{dc}|$, $j^{ac}_y=j^a\cos\alpha$, $j^{ac}_x=j^a\sin\alpha$, $j^a=|\mathbf{j}^{ac}|$, and $j_c\equiv cU_pk/2\Phi_0$. In equation~\eqref{e04TAvpx}
\begin{equation}
    \label{e04TAmom}
    \begin{array}{ll}
    \langle \sin \textsl{x} \rangle(t)=i[\langle r \rangle(t)-\langle r^{-1} \rangle(t)]/2,\\
    \\
    \langle \cos 2\textsl{x} \rangle(t)=[\langle r^2 \rangle(t)+\langle r^{-2} \rangle(t)]/2,
    \end{array}
\end{equation}
where $r^m(t)=\exp\{-im\textsl{x}(t)\}$ are the moments and their averages can be derived in terms of a matrix continued fraction, as detailed next.

\subsection{Solution in terms of a matrix continued fraction}
The mathematical aspects of solving the Langevin equation~\eqref{e04TAleq} by the matrix continued fraction method are comprehensively addressed in~\cite{Cof04boo}. This method has been previously used by two of us to solve the Langevin equation for a cosine WPP in~\cite{Shk08prb}. In this subsection, we extend that method to the ratchet WPP given by equation~\eqref{e04TArpp}.

We are only concerned with the \emph{stationary} ac response independent of the initial conditions. This is why, having substituted equation~\eqref{e04TArpp} into equation~\eqref{e04TAleq}, we perform the following series of transformations~\cite{Shk08prb} for the dimensionless vortex coordinate
\begin{equation}
    \label{e04TAftr}
    \textsl{x}(t)\rightarrow r^m(t)=e^{-im\textsl{x}(t)}=\sum\limits_{k=-\infty}^{\infty}F_k^m(\omega)e^{ik\omega t},
\end{equation}
where $F_k^m(\omega)$ are the Fourier amplitudes. Then, equation~\eqref{e04TAleq} can be reduced to the following recurrence equation
\begin{equation}
    \label{e04TAreq}
     \begin {array}{ll}
    i[F_k^{m-1}(\omega) - F_k^{m+1}(\omega)]  + j^{ac}[F_{k+1}^m(\omega)+F_{k-1}^m(\omega)]+\\
    \\ \qquad+ z_{m,k}(\omega)F_k^m(\omega)+ e[F_k^{m-2}(\omega) + F_k^{m+2}(\omega)] =0,
    \end{array}
\end{equation}
where
\begin{equation}
    \label{e04TAzmk}
    z_{m,k}(\omega)=2(j^{dc}+ \omega \hat{\tau} k/m -im/g).
\end{equation}

The recurrence equation~\eqref{e04TAreq} can be solved in terms of a matrix continued fraction. For this, one introduces the infinite column vectors $\mathbf{C}_m(\omega)$ containing all Fourier amplitudes $F_k^m(\omega)$, viz.,
\begin{equation}
    \label{e04TAvcm}
        \mathbf{C}_{m=\pm1,\pm2, ...}(\omega)=
        \left(
            \begin{array}{ccc}
                \vdots \\
                F^m_{-2}(\omega)\\
                F^m_{-1}(\omega)\\
                F^m_{0}(\omega)\\
                F^m_{1}(\omega)\\
                F^m_{2}(\omega)\\
                \vdots
            \end{array}
        \right)
        \hspace{3ex}\textrm{and}\hspace{3ex}
        \mathbf{C}_{m=0}=
        \left(
            \begin{array}{ccc}
                \vdots \\
                0\\
                0\\
                1\\
                0\\
                0\\
                \vdots
            \end{array}
        \right).
\end{equation}
Then, the scalar seven-term recurrence equation~\eqref{e04TAreq} can be transformed into the matrix five-term recurrence relation
\begin{equation}
    \label{e04TAm5t}
    \mathbf{Q}_m\mathbf{C}_{m} = i (\mathbf{C}_{m+1} - \mathbf{C}_{m-1}) - e(\mathbf{C}_{m+2} + \mathbf{C}_{m-2}),
\end{equation}
where $\mathbf{Q}_m\equiv\mathbf{Q}_m(\omega)$ are matrices of infinite dimension.

In order to reduce equation~\eqref{e04TAm5t} to the canonical form, one performs the changes $m\rightarrow2m-1$ and $m\rightarrow2m$, and introduces a new unknown column vector
\begin{equation}
    \label{e04TA2cm}
          \mathbf{C}_m(\omega)\rightarrow
          \left(
            \begin{array}{ccc}
                \mathbf{C}_{2m-1}\\
                \mathbf{C}_{2m}\\
            \end{array}
          \right)
          \equiv\mathbf{A}_m(\omega),
\end{equation}
that allows one to rewrite equation~\eqref{e04TAm5t} as the following three-term matrix equation
\begin{equation}
    \label{e04TAm3t}
    \mathbf{Q}_m^-\mathbf{A}_{m-1} + \mathbf{Q}_m\mathbf{A}_{m} + \mathbf{Q}_m^+\mathbf{A}_{m+1}=0.
\end{equation}
Here the three known matrices $\mathbf{Q}_m$, $\mathbf{Q}_m^+$, and $\mathbf{Q}_m^-$ are defined as
\begin{equation}
    \label{e04TA3mq}
    \begin{array}{c}
        \mathbf{Q}_m\equiv
        \left(
            \begin{array}{ccc}
               \mathbf{Q}_{2m-1} & -i\mathbf{I} \\
                i\mathbf{I} & \mathbf{Q}_{2m} \\
            \end{array}
        \right),\\
       \mathbf{Q}_m^+\equiv
       \left(
            \begin{array}{ccc}
                e\mathbf{I} & 0 \\
                -i\mathbf{I} & e\mathbf{I} \\
            \end{array}
       \right),
       \qquad\mathrm{and}\qquad
            \mathbf{Q}_m^-\equiv
       \left(
             \begin{array}{ccc}
                e\mathbf{I} & i\mathbf{I} \\
                0 & e\mathbf{I} \\
            \end{array}
       \right)
    \end{array}
\end{equation}
In equation~\eqref{e04TA3mq} $\mathbf{I}$ is the identity matrix of infinite dimension. Also the sub-matrices in $\mathbf{Q}_{m}\equiv\mathbf{Q}_{m}(\omega)$ are infinite and three-diagonal. They are given by
\begin{equation}
    \label{e04TAimq}
    \mathbf{Q}_{m}=i
     \left(
            \begin{array}{ccccc}
              \vdots    &   \vdots        &       \vdots    &   \vdots          & \vdots    \\
              \vdots    &   j^{ac}        &          0      &   \vdots          & \vdots    \\
              \vdots    &z_{m,-1}(\omega) &       j^{ac}    &   0               & \vdots    \\
              \vdots    &   j^{ac}        & z_{m,0}(\omega) &   j^{ac}          & \vdots    \\
              \vdots    &        0        &       j^{ac}    &   z_{m,1}(\omega) & \vdots    \\
              \vdots    &    \vdots       &         0       &   j^{ac}          & \vdots    \\
              \vdots    &    \vdots       &       \vdots    &   \vdots          & \vdots
        \end{array}
       \right).
\end{equation}

To solve equation~\eqref{e04TAm3t} with respect to $\mathbf{A}_m(\omega)$, we use the standard ansatz~\cite{Cof04boo}
$\mathbf{A}_m=\mathbf{S}_m\mathbf{A}_{m-1}$. It reduces equation~\eqref{e04TAm3t} to
\begin{equation}
    \label{e04TAsms}
    \mathbf{S}_m = - (\mathbf{Q}_m + \mathbf{Q}_m^+\mathbf{C}_{m+1})^{-1}\mathbf{Q}_m^-.
\end{equation}
Having substituted $\mathbf{S}_{m+1}$, $\mathbf{S}_{m+2}$, $\dots$ into equation~\eqref{e04TAsms} one obtains the solution in terms of a matrix continued fraction
\begin{equation}
    \label{e04TAmcf}
    \mathbf{S}_1=
    \frac{-\mathbf{I}}{\mathbf{Q}_1+\mathbf{Q}_1^+\displaystyle\frac{-\mathbf{I}}
    {\mathbf{Q}_2+\mathbf{Q}_2^+\displaystyle\frac{-\mathbf{I}}{\mathbf{Q}_3+
    \mathbf{Q}_3^+\displaystyle\frac{-\mathbf{I}}{...}}\mathbf{Q}_3^-}\mathbf{Q}_2^-}\mathbf{Q}_1^-~,
\end{equation}
where the fraction lines designate the matrix inversions.

Once $\mathbf{S}_1$ has been found, all the column vectors $\mathbf{C}_1$, $\mathbf{C}_{-1}$, $\mathbf{C}_2$, and $\mathbf{C}_{-2}$
can been obtained. To accomplish this, due to the relation $\mathbf{A}_1 = \mathbf{S}_1\mathbf{A}_0$ one can write
\begin{equation}
    \label{e04TAc12}
    \left(
            \begin{array}{ccc}
               \mathbf{C}_1\\
                \mathbf{C}_2
            \end{array}
    \right)
    =
       \mathbf{S}_{1}
    \left(
            \begin{array}{ccc}
                \mathbf{C}_{-1}\\
                \mathbf{C}_0
            \end{array}
    \right)
    \equiv
    \left(
            \begin{array}{ccc}
                 \mathbf{S}_{11} & \mathbf{S}_{12}\\
                    \mathbf{S}_{21} & \mathbf{S}_{22}
            \end{array}
    \right)
   \left(
            \begin{array}{ccc}
                \mathbf{C}_{-1}\\
       \mathbf{C}_0
            \end{array}
    \right),
\end{equation}
where $\mathbf{S}_{ik}$ ($i,k=1,2$) are matrices of infinite dimension.

Turning back to the Fourier amplitudes $F_k^m(\omega)$ it can be shown that these satisfy the relations
\begin{equation}
    \label{e04TAfre}
    F_0^1(\omega)=F_0^{-1\ast}(\omega) \quad \textrm{and}\quad F_k^{-1}(\omega)=F_{-k}^{1\ast}(\omega),
\end{equation}
where the asterisk denotes the complex conjugate. Accordingly, from relations~\eqref{e04TAfre} and the definition of $\mathbf{C}_m(\omega)$ by equation~\eqref{e04TAvcm} it follows that
\begin{equation}
    \label{e04TAcre}
    \mathbf{C}^*_{-m}=
    \left(
            \begin{array}{ccc}
                 \vdots \\
                (F^{-m}_k)^*\\
                \vdots
            \end{array}
    \right)
    =
    \left(
            \begin{array}{ccc}
                 \vdots \\
                (F^m_{-k})\\
                \vdots
            \end{array}
    \right)
    = \mathbf{T}\mathbf{C}_m,
\end{equation}
where $ \mathbf{T}$ is the matrix of column transposition defined by
\begin{equation}
    \label{e04TAtrm}
    \mathbf{T}
        \left(
            \begin{array}{ccc}
                 \vdots \\
                F^m_{-1}\\
                F^m_{0}\\
                F^m_{1}\\
                \vdots
            \end{array}
        \right)
        =
        \left(
            \begin{array}{ccccc}
            \vdots & \vdots & \vdots & \vdots & \vdots\\
            \vdots & 0 & 0 &1 & \vdots\\
            \vdots & 0 & 1 & 0 & \vdots\\
            \vdots & 1 & 0 & 0 & \vdots\\
            \vdots & \vdots & \vdots & \vdots & \vdots\\
            \end{array}
        \right)
        \left(
            \begin{array}{ccc}
                 \vdots \\
                F^m_{-1}\\
                F^m_{0}\\
                F^m_{1}\\
                \vdots
            \end{array}
        \right)
        =
         \left(
            \begin{array}{ccc}
                 \vdots \\
                F^m_{1}\\
                F^m_{0}\\
                F^m_{-1}\\
                \vdots
            \end{array}
        \right).
\end{equation}

Finally, taking into account equations~\eqref{e04TAcre} and~\eqref{e04TAtrm}, equation~\eqref{e04TAc12} can be solved with respect to
$\mathbf{C}_1$, $\mathbf{C}_{-1}$, $\mathbf{C}_2$, and $\mathbf{C}_{-2}$. The result is
\begin{equation}
    \label{e04TAc1s}
    \begin{array}{ll}
    \mathbf{C}_1 =\overline{(\mathbf{I} - \mathbf{S}_{11}\mathbf{T}\mathbf{S}_{11}^*\mathbf{T}})
    (\mathbf{S}_{12} + \mathbf{S}_{11}\mathbf{T}\mathbf{S}_{12}^*)\mathbf{C}_0,\\
    \mathbf{C}_{-1}=\mathbf{T}\mathbf{C}_1^*,
    \end{array}
\end{equation}
\begin{equation}
    \label{e04TAc2s}
    \begin{array}{ll}
    \mathbf{C}_2 = \mathbf{S}_{21}\overline{\mathbf{S}}_{11}(\mathbf{C}_1 - \mathbf{S}_{12}\mathbf{C}_0) + \mathbf{S}_{22}\mathbf{C}_0,\\
    \mathbf{C}_{-2}=\mathbf{T}\mathbf{C}_2^*,
    \end{array}
\end{equation}
where the overlines designate the matrix inversions.

In this way, with the help of equations~\eqref{e04TAftr},\eqref{e04TAvcm},\eqref{e04TAc1s}, and \eqref{e04TAc2s} the moments $\langle r \rangle(t)$, $\langle r^{-1} \rangle(t)$, $\langle r^2 \rangle(t)$, $\langle r^{-2} \rangle(t)$, and the corresponding Fourier amplitudes $F_k^m(\omega)$ are now known.

\subsection{Calculation of the average pinning force}
According to equation~\eqref{e04TApif} the dimensionless average pinning force is $\langle \hat{F}_{px}\rangle (t)=-\langle\sin\textsl{x}\rangle (t) + e\langle \cos2\textsl{x}\rangle (t)$. This quantity is the main anisotropic nonlinear component of the theory under discussion. The nonlinearity of $\langle \hat{F}_{px}\rangle (t)$ is due to its dependence on the ac and dc current inputs and temperature. For the subsequent analysis, taking into account equations~\eqref{e04TAmom} and~\eqref{e04TAftr}, it is convenient to expand $ \langle \hat{F}_{px}\rangle (t)$ into three parts
\begin{equation}
    \label{e04TAf3p}
    \langle \hat{F}_{px}\rangle (t)=\langle \hat{F}_{px}\rangle_0^{\omega}+ \langle \hat{F}_{px}\rangle_{t1}+\langle \hat{F}_{px}\rangle_t^{k>1},
\end{equation}
whose physical meaning is explained next.

The first term $\langle \hat{F}_{px}\rangle_0^{\omega}$ is the time-independent (but frequency-dependent) \textit{static} average pinning force. It has the form
\begin{equation}
    \label{e04TAf1t}
    \langle \hat{F}_{px}\rangle_0^{\omega}\equiv - \langle\sin\textsl{x}\rangle_0^{\omega} + e\langle\cos\textsl{2x}\rangle_0^{\omega} =
    \textrm{Im}\psi_0^{(1)} + e\textrm{Re}\psi_0^{(2)}.
\end{equation}
This term will be used for the derivation of the dc voltage response. In equation~\eqref{e04TAf1t}, $\psi_0^{(1)}\equiv F_0^1(\omega)$ and $\psi_0^{(2)}\equiv F_0^2(\omega)$. Both these are determined by equation~\eqref{e04TAvcm}.

The second term $\langle \hat{F}_{px}\rangle_{t1}$ is the time-dependent \textit{dynamic} average pinning force with the frequency $\omega$ of the ac current input. This component is responsible for the nonlinear impedance $Z_1(\omega)$ and can be expressed as
\begin{equation}
    \label{e04TAf2t}
    \langle \hat{F}_{px}\rangle_{t1}\equiv -\langle\sin\textsl{x}\rangle_{t1} + e\langle\cos\textsl{2x}\rangle_{t1}=\textrm{Im}(\psi_1^{(1)}e^{i\omega t}) + e\textrm{Re}(\psi_1^{(2)}e^{i\omega t}),
\end{equation}
where $\psi_k^{(1)}\equiv F_k^1(\omega)-F_k^{-1}(\omega)$ and $\psi_k^{(2)}\equiv F_k^2(\omega)+F_k^{-2}(\omega)$ can be found by equations~\eqref{e04TAvcm},~\eqref{e04TAc1s}, and ~\eqref{e04TAc2s}.

The third term $\langle \hat{F}_{px}\rangle_{t1}^{k>1}$ describes the contribution of \textit{higher harmonics} with $k>1$ to the dynamic average pinning force. It is determined by
\begin{equation}
    \label{e04TAf3t}
    \begin{array}{ll}
    \langle \hat{F}_{px}\rangle_{t}^{k>1}\equiv -\langle\sin\textsl{x}\rangle_{t}^{k>1} + e\langle\cos\textsl{2x}\rangle_{t}^{k>1} =\\
    \\
     \qquad\qquad\quad=\sum\limits_{k=2}^{\infty}\{\textrm{Im}(\psi_k^{(1)} e^{ik\omega t})  +  e\textrm{Re}(\psi_k^{(2)} e^{ik\omega t} )\}.
    \end{array}
\end{equation}

\subsection{Expressions for experimentally accessible values}
The main quantity of physical interest in our problem is the average electric field $\langle \mathbf{E}(t) \rangle$ induced by the vortex ensemble on move. It is given by
\begin{equation}
    \label{e04TAe0w}
    \langle \mathbf{E}(t) \rangle=(n/c)\mathbf{B}\times \langle \mathbf{v} \rangle=n(B/c)(-\langle v_{y}\rangle\mathbf{x}+\langle v_{x}\rangle\mathbf{y}),
\end{equation}
where $\mathbf{x}$ and $\mathbf{y}$ are the unit vectors in the $x$ and $y$ directions, respectively. Then, substituting equation~\eqref{e04TAvpx}
into equation~\eqref{e04TAe0w} the time-independent dc components $\langle E^{dc}_x\rangle_0^{\omega}$ and $\langle E^{dc}_y\rangle_0^{\omega}$ can
be expressed as
\begin{equation}
\left\{
    \begin{array}{ll}
    \label{e04TAedc}
    \langle E^{dc}_y\rangle_0^{\omega}= \rho_f\nu_0^{\omega}(j^{dc}_y + \delta j_x^{dc})/D,\\
    \\
    \langle E^{dc}_x\rangle_0^{\omega}= \rho_fj^{dc}_x - \delta\langle E^{dc}_y\rangle_0^{\omega},\\
    \end{array}
    \right.
\end{equation}
where $\rho_f = B\Phi_0/\eta c^2$ is the flux-flow resistivity. Here $\nu_0^{\omega}$ is the $(j^{dc}, j^{ac}, \omega, T)$-dependent effective mobility of the vortex under the influence of the dimensionless generalized moving force $\hat{F}_{Lx}^{dc} = j^{dc}$ in the $x$ direction being
\begin{equation}
    \label{e04TAn0w}
    \nu_0^{\omega}\equiv1-[\langle \sin \textsl{x}\rangle_0^{\omega} - e\langle \cos \textsl{2x}\rangle_0^{\omega}]/j^{dc}
    =1+\langle \hat{F}_{px}\rangle_0^{\omega}/j^{dc}.
\end{equation}

The time-dependent stationary ac response is determined as
\begin{equation}
    \label{e04TAetw}
    \langle \mathbf{E} \rangle _t\equiv\langle\langle\mathbf{E} \rangle (t)-\langle\mathbf{E}\rangle_0^{\omega}\rangle=(nB/c)[\langle v_x \rangle_t\mathbf{y}-\langle v_y \rangle_t \mathbf{x}],
\end{equation}
where $\langle \mathbf{E}\rangle_0^{\omega}$ is the time-independent part of $\langle \mathbf{E} \rangle (t)$. Note that $\langle v_y \rangle_t$ and $\langle v_x \rangle_t$ are time-dependent periodic parts of $\langle v_y \rangle (t)$ and $\langle v_x \rangle (t)$ vanishing after averaging over the period $2\pi/\omega$ of an ac cycle in equation~\eqref{e04TAedc}. Then, from equations~\eqref{e04TAedc} and~\eqref{e04TAetw}
    \begin{equation}
        \label{e04TAeac}
        \left\{
        \begin{array}{ll}
        \langle E_y^{ac}\rangle_t = (n\rho_fj_c/D) \sum_{k=1}^{\infty}(j^{ac})^k \textrm{Re}\{Z_k(\omega)e^{ik\omega t}\},\\
        \\
        \langle E_x^{ac}\rangle_t = \rho_f j_x^{ac}\cos\omega t - \delta \langle E_y^{ac}\rangle_t,
        \end{array}
        \right.
    \end{equation}
with
    \begin{equation}
        \label{e04TAzkw}
        Z_k(\omega)=\delta_{1,k}- [i\psi_k^{(1)}(\omega) - e\psi_k^{(2)}(\omega)]/(j^{ac})^k,
    \end{equation}
where $\delta_{1,k}$ is Kronecker's delta. The dimensionless transformation coefficients $Z_k(\omega)$ have the physical meaning of the $k$th
harmonic with frequency $\omega_k\equiv k\omega$ in the ac nonlinear $\langle E_y^{ac}\rangle_t$ response.

The nonlinear power absorption in the ac response per unit volume and averaged over the period of an ac cycle is given in accordance with
equation~(84) of~\cite{Shk08prb} by the following expression
\begin{equation}
    \label{e04TAnpa}
    \bar{\cal{P}}(\Omega)= (\rho_f/2D)(j^{ac})^2[D\sin^2\alpha + (1-\sin^2\alpha)\textrm{Re}Z_1(\omega)],
\end{equation}
where
\begin{equation}
    \label{e04TAz1w}
    Z_1(\omega)=1-[\langle \sin\textsl{x}\rangle_{t1} - e\langle \cos\textsl{2x}\rangle_{t1}]/j^{ac} =
    1+\langle \hat{F}_{px}\rangle_{t1}/j^{ac}.
\end{equation}
is the nonlinear impedance.

To summarize, the experimentally accessible values being the dc voltage, the contribution of the $k\omega$-harmonics in the ac response, and the absorbed ac power are determined by equations~\eqref{e04TAedc},~\eqref{e04TAeac}, and~\eqref{e04TAnpa}, respectively. Their behavior as functions of all driving parameters in the problem is analyzed next.

\section{Graphical analysis}
\label{c04TAgra}
Our objective in this section is to analyze the dc voltage response and the absorbed ac power as functions of their driving parameters. These are the dimensionless dc bias $\xi^d=j^d/j_c$, the dimensionless amplitude $\xi^a=j^a/j_c$ and the dimensionless frequency $\Omega=\omega\hat{\tau}$ of the ac input. An analysis of the response on the $k\omega$-frequency will be reported elsewhere. It is worth noting that equations~\eqref{e04TAedc} and equation~\eqref{e04TAnpa} are written for arbitrary values of $\alpha$ and $\epsilon$. However, to simplify the interpretation of the obtained results, below the Hall effect will be neglected ($\epsilon=0$) and we will put emphasis on the case when $\alpha=0^{\circ}$. In this geometry both currents flow along the WPP channels provoking the vortex movement perpendicular to them. As a result, below we consider only the nonlinear $y$ component of the dc voltage response. We will omit all indices and $\langle\dots\rangle$ in $E$ given by equation~\eqref{e04TAedc} to simplify the notation.

All data presented in the following figures are calculated for the dimensionless inverse temperature $g\equiv U_p/2T=100$. This represents a reasonable, experimentally achievable value, e.g., for thin Nb films either grown on facetted sapphire substrates~\cite{Sor07prb,Hut02afm,Ost05jap} or furnished with nano-fabricated WPP landscapes~\cite{Dob12njp}, where $U_p\simeq 5000$~K and $T\approx8$~K. The frequency $\Omega$ is measured in units of the depinning frequency $\omega_p=1/\hat{\tau}$. This frequency $\omega_p$ determines the transition from the weakly dissipative to strongly dissipative regimes in the vortex dynamics~\cite{Shk12inb}. For unpatterned films, the experimentally deduced value of the depinning frequency in the absence of a dc current at a temperature of $\simeq0.6T_c$ is $\omega_p\approx7$~GHz for 20~nm-thick~\cite{Pom10pcs} and 40~nm-thick~\cite{Jan06prb} Nb films. For comparison, $\omega_p<15$~GHz in YBCO, typically~\cite{Sil06inb}.

Before entering the discussion, the actual calculation procedure relying upon the exact matrix continued fraction solution~(\ref{e04TAmcf}) should be commented on. In the calculation, the infinite matrix continued fraction in equation~(\ref{e04TAmcf}) was approximated by a matrix continued fraction of finite order. This has been done by putting $\mathbf{Q}_m = \mathbf{0}$ at some $m = M$, whereas the dimension of the submatrices $\mathbf{Q}_m$ and the vectors $\mathbf{C}_m$ was confined to some finite number $K$. Both $M$ and $K$ depend on the parameters $g$ and $\xi^a$ and on the number of harmonics to be taken into account. These numbers were chosen as $K = 61$ and $M=200-1000$ for the reliable calculation of the components $F^1_k(\omega)$ for $k$ up to 10, for $\xi^a$ up to $10$, and for $g=100-1000$, respectively. This has ensured a calculation accuracy of not less than four digits for the majority of cases.
\begin{figure}
    \centering
    \includegraphics[width=0.4\textwidth]{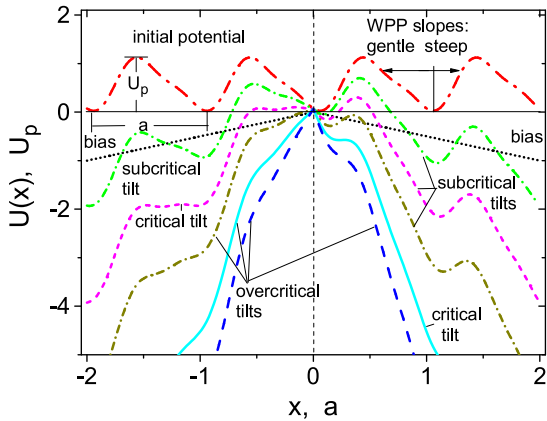}
    \caption{Modification of the effective ratchet WPP $U(x)\equiv U_p(x) - Fx$ with gradual increase of the Lorentz force component in the $x$-direction $F$. The data are plotted for $e=0.5$ for the ratchet WPP given by equation~\eqref{e04TArpp}. The motive force $F$ is applied against the steep-slope ($F_{p~steep}$ for $x>0$) and the gentle-slope ($F_{p~gentle}$ for  $x<0$) WPP directions. Accordingly, depending on the bias value, in the absence of an ac current and assuming $T=0$ for simplicity, the vortex movement in a tilted ratchet WPP has the following regimes: (i) If $F < F_{p~gentle}, F_{p~steep}$ the vortex is in the localized state. With further increase of the bias value the gentle-slope barrier vanishes, i.e., the critical tilt is achieved with respect to the left WPP barrier. (ii) If $F_{p~gentle} < F < F_{p~steep}$, the running mode in the vortex motion appears in the gentle-slope direction of the WPP. Whereas the vortex remains in the localized state with respect to the steep-slope direction of the WPP. Further increase of the bias value leads to the vanishing of the right WPP barrier, i.e., the steep-slope critical tilt is achieved. (iii) When $F_{p~gentle}, F_{p~steep} < F$, the running state in the vortex motion is realized with respect to both, gentle- and steep-slope directions of the WPP.}
   \label{f04TAtrp}
\end{figure}
Two groups of new results caused by the competition between the internal and the external anisotropy of the WPP are reported next.

\subsection{Electric field dc response}
The ratchet current-voltage curve (CVC) $E(\xi^a)$ has been thoroughly analyzed in the adiabatic, intermediate-, and high-frequency regimes for a cosine WPP in~\cite{Shk11prb}. Here we will discuss only those distinct features originating from the asymmetry of the WPP. The central notion in this discussion will be the critical tilt, as is explained in the caption of figure~\ref{f04TAtrp}. If one defines the critical dc bias $\xi^d_c$ as that corresponding to the vanishing WPP barrier in figure~\ref{f04TAtrp}, then it is evident that for the gentle-slope (left) WPP barrier the critical dc density $|-\xi^d_{c~gentle}|$ is less than $\xi^d_{c~steep}$ for the steep-slope (right) WPP barrier. This feature allows for the rectification of ac signals (the diode effect), provided the ac amplitude satisfies the condition $|-\xi_{c~gentle}|\leq\xi^a\leq\xi_{c~steep}$.

\begin{figure*}
    \centering
    \includegraphics[width=0.8\textwidth]{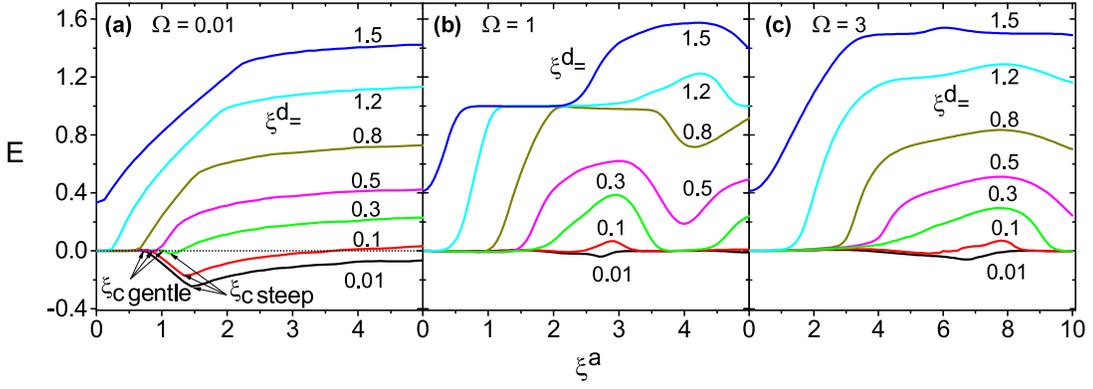}
    \caption[Dc-voltage-ac-drive response]
            {The ratchet voltage $E(\xi^a)$ for a set of dc biases $\xi^d$, as indicated, in the adiabatic~(a), intermediate-~(b), and high-frequency~(c) regime.}
  \label{f04TAeva}
\end{figure*}

The dc ratchet voltage $E(\xi^a)$ is plotted for a set of dc densities $\xi^d$ in figure~\ref{f04TAeva}. Consider first the adiabatic case ($\Omega=0.01$) in figure~\ref{f04TAeva}(a). Depending on the tilting bias value, one can distinguish three qualitatively different regimes in $E(\xi^a)$. For $\xi^d \lesssim 0.4$, with increasing $\xi^a$ the curves demonstrate a zero plateau for $\xi^a < \xi^{a}_{c~gentle}\simeq0.6$ followed by a negative voltage having a minimum at $\xi^a_{c~steep}$ and a sign change, i.e., the effect of ratchet reversal. By contrast to this scenario, for $\xi^d \gtrsim0.4$ the bias-induced asymmetry prevails over the internal asymmetry of the WPP irregardless of $\xi^a$. As a result, the overcritical tilts for the gentle-slope WPP direction are not achievable during the negative portion of the ac period and thus, the running vortex state in the gentle-slope (negative) direction of the $x$ axis is not possible. For this reason $E(\xi^a)$ is non-negative in the entire parameter range. At the same time, a zero plateau in $E(\xi^a)$ is maintained as long as the steep-slope critical current is not reached. Lastly, for $\xi^d>\xi^d_{c~steep}\simeq1.4$ the running vortex state is realized irrespective of $\xi^a$.
\begin{figure*}
    \centering
    \includegraphics[width=0.8\textwidth]{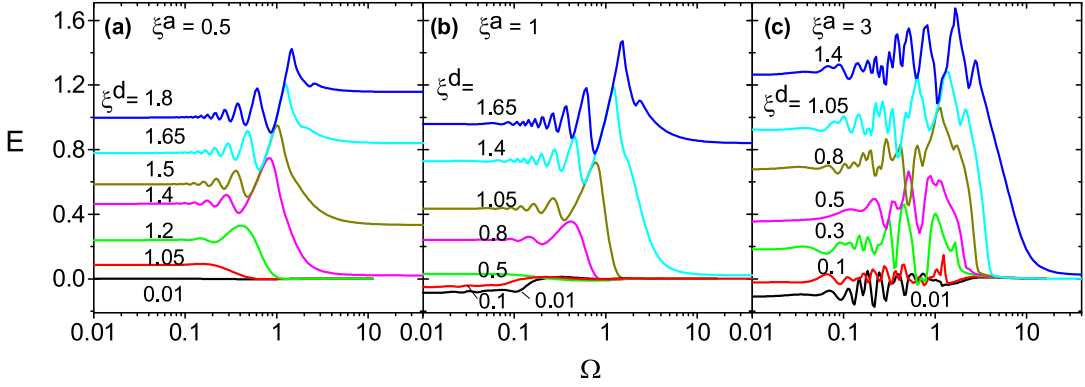}
    \caption[DC-voltage as a function of frequency]
            {The ratchet voltage $E$  versus $\Omega$ for a set of dc biases $\xi^d$, as indicated, at small (a), intermediate (b), and strong (c) ac drives.}
   \label{f04TAevw}
\end{figure*}

Turning now to the intermediate- and high-frequency regimes in figures~\ref{f04TAeva}(b) and~\ref{f04TAeva}(c) we notice that the curves also largely exhibit the phase-locking features discussed previously in~\cite{Shk11prb}. The voltage reversal effect persists at very small tilting biases only while the effect itself is very small. This is due to the fact that with increasing frequency the vortices tend to oscillate within one and the same WPP well. This corresponds to the localized vortex state~\cite{Shk08prb}. Due to this depinning scenario, the impact of the pinning potential on the vortex dynamics vanishes at very high ac frequencies. Accordingly, the vortex motion becomes less sensitive to the asymmetry of the WPP as the potential itself comes out of play. As a result, intermediate and high frequencies $\Omega >1$ are unfavorable for observing the switching between the direct and reversed net motion in our problem.

The frequency dependence of the dc voltage is shown in figure~\ref{f04TAevw}. It is evident from figure~\ref{f04TAevw}(a) that the average voltage response is zero as long as the critical tilts of the WPP are not achieved by the driving stimuli in the weak-drive regime $\xi^d \lesssim 1$. For $1\lesssim\xi^d \lesssim 1.4$ the curves start at a constant value, then develop a phase-locking fringe followed by a maximum and finally approach zero. The stronger the dc bias the higher the dissipation. For overcritical dc biases $\xi^d \gtrsim 1.4$ with respect to the steep WPP barrier the curves $E(\Omega)$ saturate at constant values at $\Omega\rightarrow\infty$. Another scenario occurs for intermediate drives $\xi^a =1$ in figure~\ref{f04TAevw}(b). For $\xi^d \lesssim0.3$ the voltage is negative up to about $\Omega \simeq0.2$ and then tends to zero. For $\xi^d \gtrsim 0.3$ the behavior of the curves is qualitatively similar to that at weak ac drives and subcritical dc biases. In the regime of strong ac drives, as representative for $\xi^a = 3$ in figure~\ref{f04TAevw}(c), $E(\Omega)$ are non-regularly oscillating curves whose main value largely replicate the behavior of those at weaker ac drives and subcritical dc biases. In the regime of strong ac drives one can also recognize multiple voltage reversals ensuing at very small dc biases $\xi^d \lesssim 0.1$.

\begin{figure}
    \centering
    \includegraphics[width=0.35\textwidth]{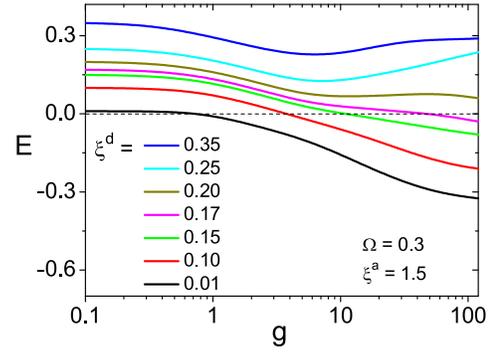}
    \caption[DC-voltage as a function of temperature]
            {The temperature dependence of the dc voltage $E(g)$ in the strong ac drive regime for a set of dc biases, as indicated.}
   \label{f04TAevg}
\end{figure}

As for the influence of the temperature $g$ on the rectified voltage $E$, it should be mentioned that increasing temperature can itself cause a sign change in $E$, as exemplified in Fig.~\ref{f04TAevg} for subcritical dc biases and overcritical ac drives. In the range $0.01\lesssim\xi^d\lesssim0.2$ and $1\lesssim g\lesssim500$ one observes the temperature-activated voltage switching effect from negative $E$ at low temperatures (large $g$) to positive $E$ at elevated temperatures. This effect can be explained as follows: An increase of temperature increases the hopping rate of the vortex for overcoming the WPP barriers, that disturbs the fragile balance between the internal and external asymmetries. At substantially high temperatures ($g\leq1$), the negative net voltage is no longer possible as the temperature-mediated transition to the direct free flux flow regime in the vortex motion ensues. In this regime, the voltage (in units of $\rho_f$) is equal to the dc current magnitude. Thus, the temperature $g$ is one additional parameter for fine-tuning the sign of the net voltage. This is owing to the strong non-linearity of the average pinning forces for the gentle-slope and steep-slope WPP directions as functions of the tilting bias $\xi^d$ and the temperature $g$.

Summing up what has been said so far, three concluding remarks should be made. Firstly, a sign change in the ratchet voltage results only for overcritical ac densities with respect to the gentle-slope WPP direction. This condition is needed for enabling the reversed rectification effect, as the gentle-slope direction of the WPP corresponds to the negative direction of the $x$ axis in our problem. Secondly, a ratchet reversal is possible only for those small dc biases for which the internal asymmetry of the WPP is not completely suppressed by the bias-induced external asymmetry. Lastly, the range of $\xi^d$ for the VR reversal effect to appear is broader in the adiabatic regime. This is because for reducing the frequency-dependent depinning the ac frequency must be small enough.

\subsection{Power absorption in ac response}
Before entering the discussion, let us recall that two physically different regimes in the vortex motion ensue depending on the driving parameters of the problem. The first regime is the localized mode when the vortex is oscillating within one WPP well. In the this regime the pinning forces dominate and the response is weakly dissipative. This regime ensues at low frequencies $\omega \ll \omega_p$ ($\Omega \ll 1$). By contrast to this, at high frequencies $\omega \gg \omega_p$ ($\Omega \gg 1$) frictional forces dominate and the response is strongly dissipative. A treatment of both regimes in terms of simple functions at $T =0$ and $0 < \xi^d < 1$ can be found, e.g., in~\cite{Shk12inb}.

\begin{figure}
    \centering
    \includegraphics[width=0.35\textwidth]{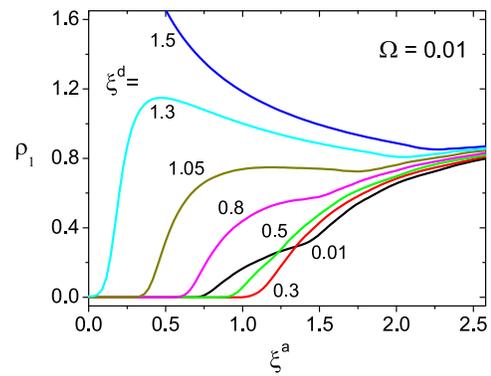}
    \caption[$\rho_1$ as a function of ac amplitude]
            {$\rho_1$  versus $\xi^a$ in the adiabatic regime for a set of dc biases $\xi^d$, as indicated.}
   \label{f04TAr1a}
\end{figure}

\begin{figure}
    \centering
    \includegraphics[width=0.35\textwidth]{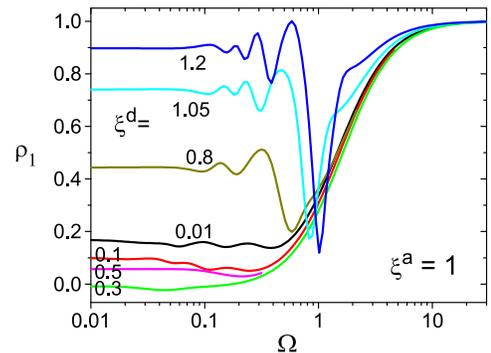}
    \caption[$\rho_1$ as a function of ac amplitude]
            {$\rho_1$  versus $\Omega$ at intermediate ac drives for a set of dc biases $\xi^d$, as indicated.}
   \label{f04TAr1w}
\end{figure}

The power absorbed per unit volume and averaged over the period of an ac cycle $\bar{\cal{P}}(\omega)$ can be written for $\alpha = 0^{\circ}$ in accordance with equation~\eqref{e04TAnpa} as
\begin{equation}
    \label{e04TA0pa}
    \bar{\cal{P}}(\Omega) = \rho_f\frac{(\xi^a)^2}{2} \mathrm{Re}Z_1(\Omega),
\end{equation}
where $Z_1(\xi^a,\xi^d,\Omega,g)$ is the nonlinear dc and ac amplitude-, frequency- and temperature-dependent impedance. As can be seen from equation~\eqref{e04TA0pa}, for analyzing the dependence $\bar{\cal{P}}(\xi^a,\xi^d,\Omega)$ at at $g = 100$ it is sufficient to calculate the ac resistivity $\rho_1 \equiv \mathrm{Re}Z_1(\xi^a,\xi^d,\Omega)$. The results are presented in figures~\ref{f04TAr1a} and~\ref{f04TAr1w}.

Consider at first the curves in figure~\ref{f04TAr1a} at $\Omega = 0.01$ corresponding to the adiabatic case. For subcritical tilts with respect to the steep slope of the WPP, $\xi^d \lesssim \xi^d_{c~steep}$, an absorption threshold occurs in the $\rho_1(\xi^a)$ curves. Its position coincides with the thresholds in the corresponding ratchet voltage curves in figure~\ref{f04TAeva}(a). At close-to-critical ac amplitudes with respect to the steep slope of the WPP the curves $\rho_1(\xi^a)$ demonstrate a nonlinear transition to unity. For close-to-critical dc biases $\xi^d \simeq 1.2$ a maximum is developing in the curves at subcritical ac amplitudes $\xi^a \simeq 0.7$. For over-critical biases $\xi^a \gtrsim 1.4$ an enhanced power absorption ensues already for weak ac drives. The reason for this is that the vortex running state is achieved by the overcritical tilting bias. In the case of intermediate and high frequencies (not shown) the curves $\rho_1(\xi^a,\xi^d,\Omega)$ quickly approach unity as the high-frequency drive shakes the vortex within one and the same WPP well which can be treated as the effective vanishing of the pinning.

We now turn to the analysis of the frequency dependence of $\rho_1$ shown in figure~\ref{f04TAr1w} for close-to-critical ac drives $\xi^a\simeq1$. We begin with the consideration of a special case $\xi^d = 0.3$. This curve is interesting as it is a smooth step-like function being in good agreement with the results of Coffey and Clem~\cite{Cof91prl} who calculated in linear approximation in $\xi^a$ the temperature dependence of the depinning frequency in a nontilted cosine pinning potential. We note that for $e = 0.5$ the bias $\xi^d = 0.3$ is very close to the gentle-slope critical value. Physically this means that this curve corresponds to a virtually complete ``compensation'' of the internal asymmetry by the external bias-induced asymmetry. If one now reduces the bias value, an enhanced absorption results at low frequencies due to the running vortex state owing to the rectification effect in the gentle-slope direction of the WPP. In the opposite case of stronger biases an enhancement of the absorbed power ensues thanks to the running state caused by the over-critical total (dc+ac) densities with respect to the steep-slope direction. The behavior of $\rho_1(\Omega)$ for overcritical dc biases is non-monotonic: with increasing $\Omega$ peculiarities in the curves become more pronounced and a sharp minimum develops at $\Omega\simeq1$. The appearance of this frequency- and temperature-dependent minimum was discussed in more detail in~\cite{Shk08prb}. With increasing $\xi^d$ the minimum shifts towards higher frequencies.

At strong ac drives (not shown) $\rho_1(\Omega)$ acquires large values already at very low frequencies $\Omega\ll1$ and quickly approaches unity. The response becomes almostly independent of the dc tilting bias. In the opposite limiting case of small ac drives (not shown) $\rho_1(\Omega)$ replicates the well-known Coffey-Clem results for all biases up to
$\xi^d \simeq 1.2$. Summing up, the variety of nonlinear regimes the vortex ensemble passes through is most rich for intermediate ac drives. Thus, this regime is most favorable for observing both, the non-monotonic dependence of the absorbed ac power at very low frequencies and the dc bias-dependent minimum in $\rho_1(\Omega)$ at $\Omega \simeq 1$.

Up to this point, the following conclusions can be drawn: In the limiting case of virtually ``compensated'' asymmetries the well-known results of Coffey and Clem at $\xi^a \ll 1$ are reproduced~\cite{Cof91prl}. At strong dc biases an enhanced power absorption results at low frequencies, having a deep minimum at $\Omega\simeq1$. At weak dc biases a non-monotonic dependence of $\rho_1(\xi^d,\Omega)$ on $\xi^d$ follows. From the analysis above we notice that the predicted effects mainly ensue at low frequencies $\omega\ll\omega_p$. This is due to the fact that the WPP depth and its asymmetry are virtually left intact by the depinning effects which are pronounced at elevated frequencies. Accordingly, more attention to the low-frequency regime will be paid in the next section.

\section{Discussion}\label{c04TAalc}
In this section we will consider the physically important limiting case of low frequencies at zero temperature. Our objective is to elucidate the scenario of the VR reversal to appear in the problem of question. This will allow us to supply the exact results discussed so far with a more visual and intuitive interpretation. It should be noted that the ratchet response at $T = 0$ can be calculated with the help of the conventional stationary CVC. For this reason, at zero temperature, we first consider the CVC for a cosine potential and next derive the CVC for the asymmetric potential given by equation~\eqref{e04TArpp}.

Consider the vortex motion in the symmetric WPP $U_p(x) = (U_p/2)[1-\cos kx]$ subject to an arbitrary current $j$ at zero temperature. Suppose that $j\equiv j(t)$ changes its sign very slowly with $\omega \rightarrow 0$. In this case one can perform the change $\textsl{x}(t)\rightarrow r^m = \exp\{-im\textsl{x}\}$ in the Langevin equation~\eqref{e04TAleq}. Accordingly, from equation~\eqref{e04TAprx} the equation for the moments $r^m$ reads
\begin{equation}
    \label{e04TAar2}
    i j r^m = [r^{m-1} - r^{m+1}]/2,
\end{equation}
where we have neglected the term $\hat{\tau}d(r^m)/dt$.

Having divided both its parts by $r^{m-1}$, one can rewrite equation~\eqref{e04TAar2} as
\begin{equation}
    \label{e04TAr20}
    r^2 + 2ij r - 1 = 0,
\end{equation}
where
\begin{equation}
    \label{e04TArpm}
    r_{\pm}= \left\{
        \begin{array}{l}
        -ij \pm \sqrt{1 - j^2},\qquad\qquad |j| < 1,\\
       -ij \pm i\sqrt{j^2 - 1},\qquad\qquad |j| > 1\\
        \end{array}
        \right.
 \end{equation}
are its roots.

Accordingly, in the adiabatic regime equation~\eqref{e04TAvpx} reads $\langle v_x \rangle(t) = \Phi_{0}j_c [j - \langle \sin \textsl{x}\rangle(t)]/c\eta$. Having in mind that in accordance with equation~(\ref{e04TAftr}) $\langle\sin \textsl{x}\rangle(t) = -\mathrm{Im} r$ one obtains the CVC sought for
\begin{equation}
    \label{e04TAarl}
    E_0 (j) = j + \mathrm{Im} r = \left\{
        \begin{array}{ll}
        0,\qquad\qquad\quad |j| < 1,\\
        \pm\sqrt{j^2 - 1},\quad~ |j| > 1.\\
        \end{array}
        \right.
\end{equation}
We note that equation~\eqref{e04TAarl} corresponds to the CVC of a resistively shunted Josephson junction, as detailed, e.g., in the book of Likharev~\cite{Lik86boo}.

We proceed now to the calculation of the CVC for the asymmetric potential $U_p(x) = (U_p/2)[1-\cos kx + e(1 -\sin 2kx)/2]$ at zero temperature. In this case the equation for the moments $r^m$ reads
\begin{equation}
    \label{e04TAara}
    2 i j = (r^{-1} - r) - ie(r^{-2} + r^2).
\end{equation}
Equation~\eqref{e04TAara} can be solved by making the change $y = r^{-1} - r$. Its roots are
\begin{equation}
    \label{e04TAar4}
    r_{\pm} = (iD \pm \sqrt{4 - D^2})/2,
\end{equation}
where $D \equiv A/2e = (1 - \sqrt{1 + 8e(e + j)})/2e$.
\begin{figure}
    \centering
    \includegraphics[width=0.45\textwidth]{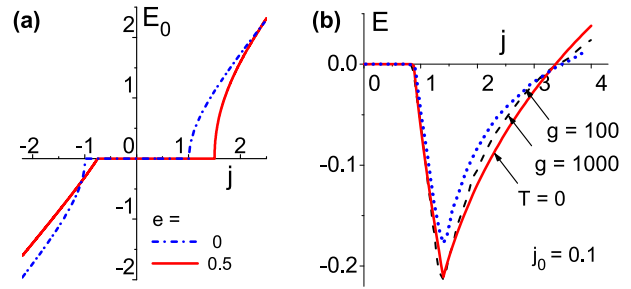}
    \caption[Origin of ratchet reversal]
            {(a)~The CVC $E_0(j)$ calculated by equation~\eqref{e04TAare} in the absence of a dc bias in the adiabatic case at zero temperature. (b)~The ratchet voltage at $T = 0$ calculated by equation~\eqref{eintegral} (solid) is plotted together with the exact solution calculated by equation~\eqref{e04TAeac} for $E^d(j)$ for $\Omega = 0.001$, $j_0 = 0.1$ at $g=1000$~(dashed) and $g=100$~(dotted).}
   \label{f04TAorr}
\end{figure}
\begin{figure*}
    \centering
    \includegraphics[width=0.69\textwidth]{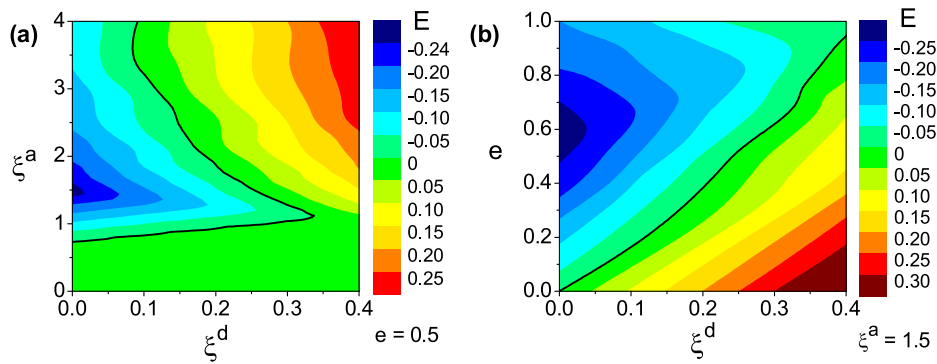}
    \caption[Ratchet reversal window]
            {Contour plots $E(\xi^d, \xi^a)$ for $e=0.5$ (a) and $E(e,\xi^d)$ for $\xi^a = 1.5$ (b) illustrate the reversal of the ratchet voltage. The thick solid lines correspond to $E = 0$.}
   \label{f04TArrw}
\end{figure*}

The CVC for the asymmetric potential is given by
\begin{equation}
    \label{e04TAare}
    E_0 (j) = j + \mathrm{Im} r + e[ 1 - 2 (\mathrm{Im} r)^2 ]
\end{equation}
and is illustrated in figure~\ref{f04TAorr}(a) for $e=0.5$ together with the CVC for the symmetric potential ($e=0$). One may notice the difference between the steep-slope and the gentle-slope critical currents due to the asymmetry of the potential. It can also be shown that the CVC $E_0(j)$ in figure~\ref{f04TAorr}(a) coincides with the dc CVC $E(\xi^d)$ calculated by equation~\eqref{e04TAedc} in the limit of zero temperature and zero ac frequency.

Our objective now is to calculate the dc ratchet response $E^r$ in the presence of a dc bias $j_0$. For this, one performs the change $j \rightarrow j\cos\omega t + j_0$ (say $j_0 = 0.1$ for definiteness). As we are seeking for the average dc ratchet voltage $E^r$ in response to the superimposed stimuli ($j\cos\omega t + j_0$), one needs to integrate the function $E$ over the ac period $T_\omega$
\begin{equation}
    \label{eintegral}
    E^r\equiv\frac{1}{T_{\omega}}\int_{0}^{T_{\omega}}dt~E(j\cos\omega t + j_0).
\end{equation}
Equation~\eqref{eintegral} can be reduced to the sum of two integrals for the positive and negative half-waves of the ac drive. We perform calculations in the spirit of~\cite{Shk11prb} that allows us to take the resulting integrals over the appropriate range of $t$. This is because of the sign-changing expressions under the radicals in equation~\eqref{e04TAar4}. The final result for the dc ratchet voltage $E^r(j\cos\omega t + j_0)$ is shown in figure~\ref{f04TAorr}(b) by the solid line. The exact solution given by equation~\eqref{e04TAedc} for the same bias value $j_0 = 0.1$ is shown by the dashed line. Both curves practically coincide: They have a zero plateau for $j \lesssim 0.7$ followed by a minimum with $E <0$. With further increasing $j$ both curves demonstrate a sign reversal, have a maximum and then tend to zero at very large $j$ (not shown).

As a generalization of the obtained results, in figure~\ref{f04TArrw} we present the entire $(\xi^d, \xi^a, e)$-parameter space at $g = 100$ where a VR reversal effect results in the adiabatic regime $\Omega = 0.01$. It should be noted that the VR reversal effect is only possible at close-to-critical ac amplitudes and deeply subcritical dc biases, as depicted in figure~\ref{f04TArrw}(a). One can trace that with increasing dc bias $\xi^d$ the gentle-slope critical current $\xi_{c~gentle}$ increases, whereas the steep-slope critical current $\xi_{c~steep}$ decreases. At $\xi^d\approx0.36$ the condition $\xi_{c~gentle} = \xi_{c~steep}$ is satisfied and for $\xi^d \gtrsim 0.36$ the (negative) rectifying effect is no longer possible. From figure~\ref{f04TArrw}(b) it follows that the reversed net motion (negative voltage) ensues in a most wide range of dc biases $0 < \xi^d \lesssim 0.4$ at $e\approx 1$. At the same time, it is evident that the anisotropy parameter $e\simeq0.5$ should be favored for observing a most sequence of the different regimes in the rectified net motion of vortices.

\section{Conclusion}\label{c04TAcon}
We have proposed an exactly solvable two-dimensional model structure for the study of the dc voltage response and the absorbed ac power in superconducting films with an asymmetric (ratchet) washboard pinning potential subject to superimposed dc and ac stimuli. We have theoretically examined the strongly nonlinear behavior of these responses as functions of the (dc+ac) transport current density $\mathbf{j}$, the frequency $\omega$, the temperature $T$, and the current tilt angle $\alpha$ with respect to the guiding direction of the WPP. It is physically clear that current, temperature, and angle $\alpha$ have qualitatively different effects on the weakening of the pinning and the corresponding transition from anisotropic vortex dynamics to isotropic. With increasing $\mathbf j$ the Lorentz force $\mathbf{F}_L$ grows and the height of the potential barrier decreases, so for $j \geqslant j_{cr1},j_{cr2}$ these barriers essentially disappear. Here $j_{cr1,2}$ are the crossover currents for these transitions to occur regarding the gentle- and steep-slope WPP barriers. In the absence of a dc bias, $j_{cr1} < j_{cr2}$ owing to the internal asymmetry of the WPP. In the presence of a dc bias, this balance is changed depending on the bias value so that the opposite inequality $j_{cr2} < j_{cr1}$ can be achieved. The intensity of the internal asymmetry defined by the parameter $e$ in equation~\eqref{e04TArpp} appears to be tunable in a narrower range as compared to the external bias-caused asymmetry for which even far overcritical tilts of the WPP can be achieved. Accordingly, the reversed net motion due to the internal asymmetry is possible at weak dc biases only, at strong dc biases the direct net transport prevails over the reversed one.

Proceeding now to a short description of the main theoretical results, we note that an exact analytical representation of the nonlinear (dc+ac)-driven responses of the investigated system in terms of a matrix continued fraction was possible thanks to the use of a simple but physically realistic model of anisotropic pinning with an asymmetric WPP. Firstly, the exact solution obtained made it possible to consistently analyze not only the qualitatively clear vortex dynamics of the ratchet reversal effect, but also non-trivial features in the conventional CVC, at any arbitrary values of the driving parameters. Secondly, the proposed model allows one to exactly calculate the nonlinear power absorbtion being usually a non-trivial issue in other models. Besides, it should be stressed that in our model the VR reversal effect unavoidably results when tuning the dc bias from zero up to overcritical values, as the two competing asymmetries having opposite effects on the vortex motion are initially included in the formulation of the problem.

With respect to the experimental verification of the theory, to the best of our knowledge, there has been no direct examination of the predicted effects so far. However, in connection with our theoretical predictions three recent studies~\cite{Jin10prb, Wor12prb} on nano-patterned superconductors should be mentioned. B.~B. Jin \emph{et al.}~\cite{Jin10prb} have experimentally investigated the frequency dependence of the dc voltage at large amplitudes of the ac driving force in the frequency range between $0.5$~MHz and $2$~GHz. A strong dependence of the ratchet reversals at frequencies higher than $3$~MHz has been reported~\cite{Jin10prb}. At lower frequencies, their results were virtually frequency-independent, as in the case of~\cite{Lar09prb}. W\"ordenweber \emph{et al.}~\cite{Wor12prb} have reported results of combined dc and microwave electronic measurements of magnetic flux transport in (sub-)micron patterned high-$T_c$ films. They have shown that the guided flux motion persists up to $8$~GHz, the highest frequency achievable in that experiment. Recently, one of us~\cite{Dob12njp} has made electronic transport measurements of the mixed state of epitaxial Nb films patterned with a symmetric pinning landscape induced by a nanogroove array. It has been shown that all anisotropic effects in the two-dimensional vortex dynamics are clearly seen. Currently under way are measurements of the magneto-resistive response of superconductor Nb films with a ratchet WPP similar to that presented in~\cite{Dob12ppa}, with an extension of ac frequencies towards the microwave range. Summing up, we consider these systems~\cite{Jin10prb,Wor12prb, Dob12njp} to be most promising for the experimental verification of the results reported here, provided the WPP in such samples is tailored in an asymmetric fashion.

It should be noted that the realization of the regime of strong dc biases involves certain difficulties in experiment. Indeed, when the critical current density $j_c$ is rather large, the realization of the strongly dissipative mode, in which the flux-flow resistivity $\rho_f$ can be measured, requires $j^{dc} \simeq j_c$. This is commonly accompanied by a non-negligible electron overheating in the sample~\cite{Shk80ltp,Bez92pcs} which changes the value of the sought $\rho_f$. We note, however, this difficulty can be overcome provided high-speed current sweeps~\cite{Leo11prb} or short-pulse measurements~\cite{Lik10prb} are employed.

Finally, we would like to note that the reported extension of the matrix continued fraction method for an asymmetric periodic potential in the presence of combined (dc+ac) stimuli may be useful for the wide class of nonlinear stochastic systems with an asymmetric potential under the action of superimposed constant and alternating drives. In particular, here we mean  charge-density-wave pinning~\cite{Gru88rmp}, the nonlinear impedance of Josephson junctions~\cite{Gol04rmp,Cof00prb}, nonlinear magnetization relaxation~\cite{Tit10prb}, stochastic resonance~\cite{Gam98rmp}, resonance activation~\cite{Bog98pre}, and noise enhanced stability~\cite{Man96prl}. We hope that some problems embracing an asymmetric potential in these research domains may benefit from the theoretical model structure presented here.

The authors are grateful to M. Huth for critical reading. OVD gratefully acknowled\-ges financial support from the German Research Foundation (DFG) through grant No.~DO~1511/2-1. This work was supported by the COST MP1201 NanoSC Action.


\begin{thebibliography}{72}%
\makeatletter
\providecommand \@ifxundefined [1]{%
 \@ifx{#1\undefined}
}%
\providecommand \@ifnum [1]{%
 \ifnum #1\expandafter \@firstoftwo
 \else \expandafter \@secondoftwo
 \fi
}%
\providecommand \@ifx [1]{%
 \ifx #1\expandafter \@firstoftwo
 \else \expandafter \@secondoftwo
 \fi
}%
\providecommand \natexlab [1]{#1}%
\providecommand \enquote  [1]{``#1''}%
\providecommand \bibnamefont  [1]{#1}%
\providecommand \bibfnamefont [1]{#1}%
\providecommand \citenamefont [1]{#1}%
\providecommand \href@noop [0]{\@secondoftwo}%
\providecommand \href [0]{\begingroup \@sanitize@url \@href}%
\providecommand \@href[1]{\@@startlink{#1}\@@href}%
\providecommand \@@href[1]{\endgroup#1\@@endlink}%
\providecommand \@sanitize@url [0]{\catcode `\\12\catcode `\$12\catcode
  `\&12\catcode `\#12\catcode `\^12\catcode `\_12\catcode `\%12\relax}%
\providecommand \@@startlink[1]{}%
\providecommand \@@endlink[0]{}%
\providecommand \url  [0]{\begingroup\@sanitize@url \@url }%
\providecommand \@url [1]{\endgroup\@href {#1}{\urlprefix }}%
\providecommand \urlprefix  [0]{URL }%
\providecommand \Eprint [0]{\href }%
\providecommand \doibase [0]{http://dx.doi.org/}%
\providecommand \selectlanguage [0]{\@gobble}%
\providecommand \bibinfo  [0]{\@secondoftwo}%
\providecommand \bibfield  [0]{\@secondoftwo}%
\providecommand \translation [1]{[#1]}%
\providecommand \BibitemOpen [0]{}%
\providecommand \bibitemStop [0]{}%
\providecommand \bibitemNoStop [0]{.\EOS\space}%
\providecommand \EOS [0]{\spacefactor3000\relax}%
\providecommand \BibitemShut  [1]{\csname bibitem#1\endcsname}%
\let\auto@bib@innerbib\@empty
%</preamble>
\bibitem [{\citenamefont {Reimann}(2002)}]{Rei02prt}%
  \BibitemOpen
  \bibfield  {author} {\bibinfo {author} {\bibfnamefont {P.}~\bibnamefont
  {Reimann}},\ }\href {\doibase 10.1016/S0370-1573(01)00081-3} {\bibfield
  {journal} {\bibinfo  {journal} {Physics Reports}\ }\textbf {\bibinfo {volume}
  {361}},\ \bibinfo {pages} {57 } (\bibinfo {year} {2002})}\BibitemShut
  {NoStop}%
\bibitem [{\citenamefont {H\"anggi}\ and\ \citenamefont
  {Marchesoni}(2009)}]{Han09rmp}%
  \BibitemOpen
  \bibfield  {author} {\bibinfo {author} {\bibfnamefont {P.}~\bibnamefont
  {H\"anggi}}\ and\ \bibinfo {author} {\bibfnamefont {F.}~\bibnamefont
  {Marchesoni}},\ }\href {\doibase 10.1103/RevModPhys.81.387} {\bibfield
  {journal} {\bibinfo  {journal} {Rev. Mod. Phys.}\ }\textbf {\bibinfo {volume}
  {81}},\ \bibinfo {pages} {387} (\bibinfo {year} {2009})}\BibitemShut
  {NoStop}%
\bibitem [{\citenamefont {Silhanek}\ \emph {et~al.}(2010)\citenamefont
  {Silhanek}, \citenamefont {Van~de Vondel},\ and\ \citenamefont
  {Moshchalkov}}]{Sil10inb}%
  \BibitemOpen
  \bibfield  {author} {\bibinfo {author} {\bibfnamefont {A.~V.}\ \bibnamefont
  {Silhanek}}, \bibinfo {author} {\bibfnamefont {J.}~\bibnamefont {Van~de
  Vondel}}, \ and\ \bibinfo {author} {\bibfnamefont {V.~V.}\ \bibnamefont
  {Moshchalkov}},\ }\enquote {\bibinfo {title} {Guided vortex motion and vortex
  ratchets in nanostructured superconductors},}\ in\ \href@noop {} {\emph
  {\bibinfo {booktitle} {Nanoscience and Engineering in Superconductivity}}}\
  (\bibinfo  {publisher} {Springer-Verlag, Berlin Heidelberg},\ \bibinfo {year}
  {2010})\ Chap.~\bibinfo {chapter} {1}, pp.\ \bibinfo {pages}
  {1--24}\BibitemShut {NoStop}%
\bibitem [{\citenamefont {Feynman}\ \emph {et~al.}(1963)\citenamefont
  {Feynman}, \citenamefont {Leighton},\ and\ \citenamefont {Sands}}]{Fey63boo}%
  \BibitemOpen
  \bibfield  {author} {\bibinfo {author} {\bibfnamefont {R.~P.}\ \bibnamefont
  {Feynman}}, \bibinfo {author} {\bibfnamefont {R.~B.}\ \bibnamefont
  {Leighton}}, \ and\ \bibinfo {author} {\bibfnamefont {M.}~\bibnamefont
  {Sands}},\ }\href@noop {} {\emph {\bibinfo {title} {The Feynman Lectures on
  Physics}}},\ Vol.\ \bibinfo {volume} {I. Reading}\ (\bibinfo  {publisher}
  {MA: Addison-Wesley},\ \bibinfo {year} {1963})\BibitemShut {NoStop}%
\bibitem [{\citenamefont {Svoboda}\ \emph {et~al.}(1993)\citenamefont
  {Svoboda}, \citenamefont {Schmidt}, \citenamefont {Schnapp},\ and\
  \citenamefont {Block}}]{Svo93nat}%
  \BibitemOpen
  \bibfield  {author} {\bibinfo {author} {\bibfnamefont {K.}~\bibnamefont
  {Svoboda}}, \bibinfo {author} {\bibfnamefont {C.~F.}\ \bibnamefont
  {Schmidt}}, \bibinfo {author} {\bibfnamefont {B.~J.}\ \bibnamefont
  {Schnapp}}, \ and\ \bibinfo {author} {\bibfnamefont {S.~M.}\ \bibnamefont
  {Block}},\ }\href {http://dx.doi.org/10.1038/365721a0} {\bibfield  {journal}
  {\bibinfo  {journal} {Nature}\ }\textbf {\bibinfo {volume} {365}},\ \bibinfo
  {pages} {721} (\bibinfo {year} {1993})}\BibitemShut {NoStop}%
\bibitem [{\citenamefont {Rousselet}\ \emph {et~al.}(1994)\citenamefont
  {Rousselet}, \citenamefont {Salome}, \citenamefont {Ajdari},\ and\
  \citenamefont {Prostt}}]{Rou94nat}%
  \BibitemOpen
  \bibfield  {author} {\bibinfo {author} {\bibfnamefont {J.}~\bibnamefont
  {Rousselet}}, \bibinfo {author} {\bibfnamefont {L.}~\bibnamefont {Salome}},
  \bibinfo {author} {\bibfnamefont {A.}~\bibnamefont {Ajdari}}, \ and\ \bibinfo
  {author} {\bibfnamefont {J.}~\bibnamefont {Prostt}},\ }\href
  {http://dx.doi.org/10.1038/370446a0} {\bibfield  {journal} {\bibinfo
  {journal} {Nature}\ }\textbf {\bibinfo {volume} {370}},\ \bibinfo {pages}
  {446} (\bibinfo {year} {1994})}\BibitemShut {NoStop}%
\bibitem [{\citenamefont {Zapata}\ \emph {et~al.}(1996)\citenamefont {Zapata},
  \citenamefont {Bartussek}, \citenamefont {Sols},\ and\ \citenamefont
  {H\"anggi}}]{Zap96prl}%
  \BibitemOpen
  \bibfield  {author} {\bibinfo {author} {\bibfnamefont {I.}~\bibnamefont
  {Zapata}}, \bibinfo {author} {\bibfnamefont {R.}~\bibnamefont {Bartussek}},
  \bibinfo {author} {\bibfnamefont {F.}~\bibnamefont {Sols}}, \ and\ \bibinfo
  {author} {\bibfnamefont {P.}~\bibnamefont {H\"anggi}},\ }\href {\doibase
  10.1103/PhysRevLett.77.2292} {\bibfield  {journal} {\bibinfo  {journal}
  {Phys. Rev. Lett.}\ }\textbf {\bibinfo {volume} {77}},\ \bibinfo {pages}
  {2292} (\bibinfo {year} {1996})}\BibitemShut {NoStop}%
\bibitem [{\citenamefont {Carapella}\ and\ \citenamefont
  {Costabile}(2001)}]{Car01prl}%
  \BibitemOpen
  \bibfield  {author} {\bibinfo {author} {\bibfnamefont {G.}~\bibnamefont
  {Carapella}}\ and\ \bibinfo {author} {\bibfnamefont {G.}~\bibnamefont
  {Costabile}},\ }\href {\doibase 10.1103/PhysRevLett.87.077002} {\bibfield
  {journal} {\bibinfo  {journal} {Phys. Rev. Lett.}\ }\textbf {\bibinfo
  {volume} {87}},\ \bibinfo {pages} {077002} (\bibinfo {year}
  {2001})}\BibitemShut {NoStop}%
\bibitem [{\citenamefont {Knufinke}\ \emph {et~al.}(2012)\citenamefont
  {Knufinke}, \citenamefont {Ilin}, \citenamefont {Siegel}, \citenamefont
  {Koelle}, \citenamefont {Kleiner},\ and\ \citenamefont
  {Goldobin}}]{Knu12pre}%
  \BibitemOpen
  \bibfield  {author} {\bibinfo {author} {\bibfnamefont {M.}~\bibnamefont
  {Knufinke}}, \bibinfo {author} {\bibfnamefont {K.}~\bibnamefont {Ilin}},
  \bibinfo {author} {\bibfnamefont {M.}~\bibnamefont {Siegel}}, \bibinfo
  {author} {\bibfnamefont {D.}~\bibnamefont {Koelle}}, \bibinfo {author}
  {\bibfnamefont {R.}~\bibnamefont {Kleiner}}, \ and\ \bibinfo {author}
  {\bibfnamefont {E.}~\bibnamefont {Goldobin}},\ }\href {\doibase
  10.1103/PhysRevE.85.011122} {\bibfield  {journal} {\bibinfo  {journal} {Phys.
  Rev. E}\ }\textbf {\bibinfo {volume} {85}},\ \bibinfo {pages} {011122}
  (\bibinfo {year} {2012})}\BibitemShut {NoStop}%
\bibitem [{\citenamefont {Gommers}\ \emph {et~al.}(2006)\citenamefont
  {Gommers}, \citenamefont {Denisov},\ and\ \citenamefont
  {Renzoni}}]{Gom06prl}%
  \BibitemOpen
  \bibfield  {author} {\bibinfo {author} {\bibfnamefont {R.}~\bibnamefont
  {Gommers}}, \bibinfo {author} {\bibfnamefont {S.}~\bibnamefont {Denisov}}, \
  and\ \bibinfo {author} {\bibfnamefont {F.}~\bibnamefont {Renzoni}},\ }\href
  {\doibase 10.1103/PhysRevLett.96.240604} {\bibfield  {journal} {\bibinfo
  {journal} {Phys. Rev. Lett.}\ }\textbf {\bibinfo {volume} {96}},\ \bibinfo
  {pages} {240604} (\bibinfo {year} {2006})}\BibitemShut {NoStop}%
\bibitem [{\citenamefont {P\'erez-Junquera}\ \emph {et~al.}(2008)\citenamefont
  {P\'erez-Junquera}, \citenamefont {Marconi}, \citenamefont {Kolton},
  \citenamefont {\'Alvarez-Prado}, \citenamefont {Souche}, \citenamefont
  {Alija}, \citenamefont {V\'elez}, \citenamefont {Anguita}, \citenamefont
  {Alameda}, \citenamefont {Mart\'in},\ and\ \citenamefont
  {Parrondo}}]{Per08prl}%
  \BibitemOpen
  \bibfield  {author} {\bibinfo {author} {\bibfnamefont {A.}~\bibnamefont
  {P\'erez-Junquera}}, \bibinfo {author} {\bibfnamefont {V.~I.}\ \bibnamefont
  {Marconi}}, \bibinfo {author} {\bibfnamefont {A.~B.}\ \bibnamefont {Kolton}},
  \bibinfo {author} {\bibfnamefont {L.~M.}\ \bibnamefont {\'Alvarez-Prado}},
  \bibinfo {author} {\bibfnamefont {Y.}~\bibnamefont {Souche}}, \bibinfo
  {author} {\bibfnamefont {A.}~\bibnamefont {Alija}}, \bibinfo {author}
  {\bibfnamefont {M.}~\bibnamefont {V\'elez}}, \bibinfo {author} {\bibfnamefont
  {J.~V.}\ \bibnamefont {Anguita}}, \bibinfo {author} {\bibfnamefont {J.~M.}\
  \bibnamefont {Alameda}}, \bibinfo {author} {\bibfnamefont {J.~I.}\
  \bibnamefont {Mart\'in}}, \ and\ \bibinfo {author} {\bibfnamefont {J.~M.~R.}\
  \bibnamefont {Parrondo}},\ }\href {\doibase 10.1103/PhysRevLett.100.037203}
  {\bibfield  {journal} {\bibinfo  {journal} {Phys. Rev. Lett.}\ }\textbf
  {\bibinfo {volume} {100}},\ \bibinfo {pages} {037203} (\bibinfo {year}
  {2008})}\BibitemShut {NoStop}%
\bibitem [{\citenamefont {Van~de Vondel}\ \emph {et~al.}(2011)\citenamefont
  {Van~de Vondel}, \citenamefont {Gladilin}, \citenamefont {Silhanek},
  \citenamefont {Gillijns}, \citenamefont {Tempere}, \citenamefont {Devreese},\
  and\ \citenamefont {Moshchalkov}}]{Van11prl}%
  \BibitemOpen
  \bibfield  {author} {\bibinfo {author} {\bibfnamefont {J.}~\bibnamefont
  {Van~de Vondel}}, \bibinfo {author} {\bibfnamefont {V.~N.}\ \bibnamefont
  {Gladilin}}, \bibinfo {author} {\bibfnamefont {A.~V.}\ \bibnamefont
  {Silhanek}}, \bibinfo {author} {\bibfnamefont {W.}~\bibnamefont {Gillijns}},
  \bibinfo {author} {\bibfnamefont {J.}~\bibnamefont {Tempere}}, \bibinfo
  {author} {\bibfnamefont {J.~T.}\ \bibnamefont {Devreese}}, \ and\ \bibinfo
  {author} {\bibfnamefont {V.~V.}\ \bibnamefont {Moshchalkov}},\ }\href
  {\doibase 10.1103/PhysRevLett.106.137003} {\bibfield  {journal} {\bibinfo
  {journal} {Phys. Rev. Lett.}\ }\textbf {\bibinfo {volume} {106}},\ \bibinfo
  {pages} {137003} (\bibinfo {year} {2011})}\BibitemShut {NoStop}%
\bibitem [{\citenamefont {Van~de Vondel}\ \emph {et~al.}(2009)\citenamefont
  {Van~de Vondel}, \citenamefont {Silhanek}, \citenamefont {Metlushko},
  \citenamefont {Vavassori}, \citenamefont {Ilic},\ and\ \citenamefont
  {Moshchalkov}}]{Von09prb}%
  \BibitemOpen
  \bibfield  {author} {\bibinfo {author} {\bibfnamefont {J.}~\bibnamefont
  {Van~de Vondel}}, \bibinfo {author} {\bibfnamefont {A.~V.}\ \bibnamefont
  {Silhanek}}, \bibinfo {author} {\bibfnamefont {V.}~\bibnamefont {Metlushko}},
  \bibinfo {author} {\bibfnamefont {P.}~\bibnamefont {Vavassori}}, \bibinfo
  {author} {\bibfnamefont {B.}~\bibnamefont {Ilic}}, \ and\ \bibinfo {author}
  {\bibfnamefont {V.~V.}\ \bibnamefont {Moshchalkov}},\ }\href {\doibase
  10.1103/PhysRevB.79.054527} {\bibfield  {journal} {\bibinfo  {journal} {Phys.
  Rev. B}\ }\textbf {\bibinfo {volume} {79}},\ \bibinfo {pages} {054527}
  (\bibinfo {year} {2009})}\BibitemShut {NoStop}%
\bibitem [{\citenamefont {Zarlenga}\ \emph {et~al.}(2009)\citenamefont
  {Zarlenga}, \citenamefont {Larrondo}, \citenamefont {Arizmendi},\ and\
  \citenamefont {Family}}]{Zar09pre}%
  \BibitemOpen
  \bibfield  {author} {\bibinfo {author} {\bibfnamefont {D.~G.}\ \bibnamefont
  {Zarlenga}}, \bibinfo {author} {\bibfnamefont {H.~A.}\ \bibnamefont
  {Larrondo}}, \bibinfo {author} {\bibfnamefont {C.~M.}\ \bibnamefont
  {Arizmendi}}, \ and\ \bibinfo {author} {\bibfnamefont {F.}~\bibnamefont
  {Family}},\ }\href {\doibase 10.1103/PhysRevE.80.011127} {\bibfield
  {journal} {\bibinfo  {journal} {Phys. Rev. E}\ }\textbf {\bibinfo {volume}
  {80}},\ \bibinfo {pages} {011127} (\bibinfo {year} {2009})}\BibitemShut
  {NoStop}%
\bibitem [{\citenamefont {Plourde}(2009)}]{Plo09tas}%
  \BibitemOpen
  \bibfield  {author} {\bibinfo {author} {\bibfnamefont {B.~L.~T.}\
  \bibnamefont {Plourde}},\ }\href {\doibase doi:10.1109/TASC.2009.2028873}
  {\bibfield  {journal} {\bibinfo  {journal} {IEEE Trans. Appl. Supercond.}\
  }\textbf {\bibinfo {volume} {19}},\ \bibinfo {pages} {3698} (\bibinfo {year}
  {2009})}\BibitemShut {NoStop}%
\bibitem [{\citenamefont {Lee}\ \emph {et~al.}(1999)\citenamefont {Lee},
  \citenamefont {Janko}, \citenamefont {Derenyi},\ and\ \citenamefont
  {Barabasi}}]{Lee99nat}%
  \BibitemOpen
  \bibfield  {author} {\bibinfo {author} {\bibfnamefont {C.-S.}\ \bibnamefont
  {Lee}}, \bibinfo {author} {\bibfnamefont {B.}~\bibnamefont {Janko}}, \bibinfo
  {author} {\bibfnamefont {I.}~\bibnamefont {Derenyi}}, \ and\ \bibinfo
  {author} {\bibfnamefont {A.-L.}\ \bibnamefont {Barabasi}},\ }\href
  {http://dx.doi.org/10.1038/22485} {\bibfield  {journal} {\bibinfo  {journal}
  {Nature}\ }\textbf {\bibinfo {volume} {400}},\ \bibinfo {pages} {337}
  (\bibinfo {year} {1999})}\BibitemShut {NoStop}%
\bibitem [{\citenamefont {Wambaugh}\ \emph {et~al.}(1999)\citenamefont
  {Wambaugh}, \citenamefont {Reichhardt}, \citenamefont {Olson}, \citenamefont
  {Marchesoni},\ and\ \citenamefont {Nori}}]{Wam99prl}%
  \BibitemOpen
  \bibfield  {author} {\bibinfo {author} {\bibfnamefont {J.~F.}\ \bibnamefont
  {Wambaugh}}, \bibinfo {author} {\bibfnamefont {C.}~\bibnamefont
  {Reichhardt}}, \bibinfo {author} {\bibfnamefont {C.~J.}\ \bibnamefont
  {Olson}}, \bibinfo {author} {\bibfnamefont {F.}~\bibnamefont {Marchesoni}}, \
  and\ \bibinfo {author} {\bibfnamefont {F.}~\bibnamefont {Nori}},\ }\href
  {\doibase 10.1103/PhysRevLett.83.5106} {\bibfield  {journal} {\bibinfo
  {journal} {Phys. Rev. Lett.}\ }\textbf {\bibinfo {volume} {83}},\ \bibinfo
  {pages} {5106} (\bibinfo {year} {1999})}\BibitemShut {NoStop}%
\bibitem [{\citenamefont {Togawa}\ \emph {et~al.}(2005)\citenamefont {Togawa},
  \citenamefont {Harada}, \citenamefont {Akashi}, \citenamefont {Kasai},
  \citenamefont {Matsuda}, \citenamefont {Nori}, \citenamefont {Maeda},\ and\
  \citenamefont {Tonomura}}]{Tog05prl}%
  \BibitemOpen
  \bibfield  {author} {\bibinfo {author} {\bibfnamefont {Y.}~\bibnamefont
  {Togawa}}, \bibinfo {author} {\bibfnamefont {K.}~\bibnamefont {Harada}},
  \bibinfo {author} {\bibfnamefont {T.}~\bibnamefont {Akashi}}, \bibinfo
  {author} {\bibfnamefont {H.}~\bibnamefont {Kasai}}, \bibinfo {author}
  {\bibfnamefont {T.}~\bibnamefont {Matsuda}}, \bibinfo {author} {\bibfnamefont
  {F.}~\bibnamefont {Nori}}, \bibinfo {author} {\bibfnamefont {A.}~\bibnamefont
  {Maeda}}, \ and\ \bibinfo {author} {\bibfnamefont {A.}~\bibnamefont
  {Tonomura}},\ }\href {\doibase 10.1103/PhysRevLett.95.087002} {\bibfield
  {journal} {\bibinfo  {journal} {Phys. Rev. Lett.}\ }\textbf {\bibinfo
  {volume} {95}},\ \bibinfo {pages} {087002} (\bibinfo {year}
  {2005})}\BibitemShut {NoStop}%
\bibitem [{\citenamefont {Menghini}\ \emph {et~al.}(2007)\citenamefont
  {Menghini}, \citenamefont {Van~de Vondel}, \citenamefont {Gheorghe},
  \citenamefont {Wijngaarden},\ and\ \citenamefont {Moshchalkov}}]{Men07prb}%
  \BibitemOpen
  \bibfield  {author} {\bibinfo {author} {\bibfnamefont {M.}~\bibnamefont
  {Menghini}}, \bibinfo {author} {\bibfnamefont {J.}~\bibnamefont {Van~de
  Vondel}}, \bibinfo {author} {\bibfnamefont {D.~G.}\ \bibnamefont {Gheorghe}},
  \bibinfo {author} {\bibfnamefont {R.~J.}\ \bibnamefont {Wijngaarden}}, \ and\
  \bibinfo {author} {\bibfnamefont {V.~V.}\ \bibnamefont {Moshchalkov}},\
  }\href {\doibase 10.1103/PhysRevB.76.184515} {\bibfield  {journal} {\bibinfo
  {journal} {Phys. Rev. B}\ }\textbf {\bibinfo {volume} {76}},\ \bibinfo
  {pages} {184515} (\bibinfo {year} {2007})}\BibitemShut {NoStop}%
\bibitem [{\citenamefont {Villegas}\ \emph {et~al.}(2003)\citenamefont
  {Villegas}, \citenamefont {Savel'ev}, \citenamefont {Nori}, \citenamefont
  {Gonzalez}, \citenamefont {Anguita}, \citenamefont {Garcia},\ and\
  \citenamefont {Vicent}}]{Vil03sci}%
  \BibitemOpen
  \bibfield  {author} {\bibinfo {author} {\bibfnamefont {J.~E.}\ \bibnamefont
  {Villegas}}, \bibinfo {author} {\bibfnamefont {S.}~\bibnamefont {Savel'ev}},
  \bibinfo {author} {\bibfnamefont {F.}~\bibnamefont {Nori}}, \bibinfo {author}
  {\bibfnamefont {E.~M.}\ \bibnamefont {Gonzalez}}, \bibinfo {author}
  {\bibfnamefont {J.~V.}\ \bibnamefont {Anguita}}, \bibinfo {author}
  {\bibfnamefont {R.}~\bibnamefont {Garcia}}, \ and\ \bibinfo {author}
  {\bibfnamefont {J.~L.}\ \bibnamefont {Vicent}},\ }\href {\doibase
  10.1126/science.1090390} {\bibfield  {journal} {\bibinfo  {journal}
  {Science}\ }\textbf {\bibinfo {volume} {302}},\ \bibinfo {pages} {1188}
  (\bibinfo {year} {2003})}\BibitemShut {NoStop}%
\bibitem [{\citenamefont {de~Souza~Silva}\ \emph {et~al.}(2006)\citenamefont
  {de~Souza~Silva}, \citenamefont {Van~de Vondel}, \citenamefont {Morelle},\
  and\ \citenamefont {Moshchalkov}}]{Sil06nat}%
  \BibitemOpen
  \bibfield  {author} {\bibinfo {author} {\bibfnamefont {C.~C.}\ \bibnamefont
  {de~Souza~Silva}}, \bibinfo {author} {\bibfnamefont {J.}~\bibnamefont {Van~de
  Vondel}}, \bibinfo {author} {\bibfnamefont {M.}~\bibnamefont {Morelle}}, \
  and\ \bibinfo {author} {\bibfnamefont {V.~V.}\ \bibnamefont {Moshchalkov}},\
  }\href@noop {} {\bibfield  {journal} {\bibinfo  {journal} {Nature}\ }\textbf
  {\bibinfo {volume} {440}},\ \bibinfo {pages} {651} (\bibinfo {year}
  {2006})}\BibitemShut {NoStop}%
\bibitem [{\citenamefont {Dinis}\ \emph
  {et~al.}(2007{\natexlab{a}})\citenamefont {Dinis}, \citenamefont
  {Gonz\'alez}, \citenamefont {Anguita}, \citenamefont {Parrondo},\ and\
  \citenamefont {Vicent}}]{Din07prb}%
  \BibitemOpen
  \bibfield  {author} {\bibinfo {author} {\bibfnamefont {L.}~\bibnamefont
  {Dinis}}, \bibinfo {author} {\bibfnamefont {E.~M.}\ \bibnamefont
  {Gonz\'alez}}, \bibinfo {author} {\bibfnamefont {J.~V.}\ \bibnamefont
  {Anguita}}, \bibinfo {author} {\bibfnamefont {J.~M.~R.}\ \bibnamefont
  {Parrondo}}, \ and\ \bibinfo {author} {\bibfnamefont {J.~L.}\ \bibnamefont
  {Vicent}},\ }\href {\doibase 10.1103/PhysRevB.76.212507} {\bibfield
  {journal} {\bibinfo  {journal} {Phys. Rev. B}\ }\textbf {\bibinfo {volume}
  {76}},\ \bibinfo {pages} {212507} (\bibinfo {year}
  {2007}{\natexlab{a}})}\BibitemShut {NoStop}%
\bibitem [{\citenamefont {Gillijns}\ \emph {et~al.}(2007)\citenamefont
  {Gillijns}, \citenamefont {Silhanek}, \citenamefont {Moshchalkov},
  \citenamefont {Reichhardt},\ and\ \citenamefont {Reichhardt}}]{Gil07prl}%
  \BibitemOpen
  \bibfield  {author} {\bibinfo {author} {\bibfnamefont {W.}~\bibnamefont
  {Gillijns}}, \bibinfo {author} {\bibfnamefont {A.~V.}\ \bibnamefont
  {Silhanek}}, \bibinfo {author} {\bibfnamefont {V.~V.}\ \bibnamefont
  {Moshchalkov}}, \bibinfo {author} {\bibfnamefont {C.~J.~O.}\ \bibnamefont
  {Reichhardt}}, \ and\ \bibinfo {author} {\bibfnamefont {C.}~\bibnamefont
  {Reichhardt}},\ }\href {\doibase 10.1103/PhysRevLett.99.247002} {\bibfield
  {journal} {\bibinfo  {journal} {Phys. Rev. Lett.}\ }\textbf {\bibinfo
  {volume} {99}},\ \bibinfo {pages} {247002} (\bibinfo {year}
  {2007})}\BibitemShut {NoStop}%
\bibitem [{\citenamefont {Dinis}\ \emph
  {et~al.}(2007{\natexlab{b}})\citenamefont {Dinis}, \citenamefont
  {Gonz\'alez}, \citenamefont {Anguita}, \citenamefont {Parrondo},\ and\
  \citenamefont {Vicent}}]{Din07njp}%
  \BibitemOpen
  \bibfield  {author} {\bibinfo {author} {\bibfnamefont {L.}~\bibnamefont
  {Dinis}}, \bibinfo {author} {\bibfnamefont {E.~M.}\ \bibnamefont
  {Gonz\'alez}}, \bibinfo {author} {\bibfnamefont {J.~V.}\ \bibnamefont
  {Anguita}}, \bibinfo {author} {\bibfnamefont {J.~M.~R.}\ \bibnamefont
  {Parrondo}}, \ and\ \bibinfo {author} {\bibfnamefont {J.~L.}\ \bibnamefont
  {Vicent}},\ }\href {http://stacks.iop.org/1367-2630/9/i=10/a=366} {\bibfield
  {journal} {\bibinfo  {journal} {New J. Phys.}\ }\textbf {\bibinfo {volume}
  {9}},\ \bibinfo {pages} {366} (\bibinfo {year}
  {2007}{\natexlab{b}})}\BibitemShut {NoStop}%
\bibitem [{\citenamefont {Jin}\ \emph {et~al.}(2010)\citenamefont {Jin},
  \citenamefont {Zhu}, \citenamefont {W\"ordenweber}, \citenamefont
  {de~Souza~Silva}, \citenamefont {Wu},\ and\ \citenamefont
  {Moshchalkov}}]{Jin10prb}%
  \BibitemOpen
  \bibfield  {author} {\bibinfo {author} {\bibfnamefont {B.~B.}\ \bibnamefont
  {Jin}}, \bibinfo {author} {\bibfnamefont {B.~Y.}\ \bibnamefont {Zhu}},
  \bibinfo {author} {\bibfnamefont {R.}~\bibnamefont {W\"ordenweber}}, \bibinfo
  {author} {\bibfnamefont {C.~C.}\ \bibnamefont {de~Souza~Silva}}, \bibinfo
  {author} {\bibfnamefont {P.~H.}\ \bibnamefont {Wu}}, \ and\ \bibinfo {author}
  {\bibfnamefont {V.~V.}\ \bibnamefont {Moshchalkov}},\ }\href {\doibase
  10.1103/PhysRevB.81.174505} {\bibfield  {journal} {\bibinfo  {journal} {Phys.
  Rev. B}\ }\textbf {\bibinfo {volume} {81}},\ \bibinfo {pages} {174505}
  (\bibinfo {year} {2010})}\BibitemShut {NoStop}%
\bibitem [{\citenamefont {Perez~de Lara}\ \emph {et~al.}(2010)\citenamefont
  {Perez~de Lara}, \citenamefont {Alija}, \citenamefont {Gonzalez},
  \citenamefont {Velez}, \citenamefont {Martin},\ and\ \citenamefont
  {Vicent}}]{Lar10prb}%
  \BibitemOpen
  \bibfield  {author} {\bibinfo {author} {\bibfnamefont {D.}~\bibnamefont
  {Perez~de Lara}}, \bibinfo {author} {\bibfnamefont {A.}~\bibnamefont
  {Alija}}, \bibinfo {author} {\bibfnamefont {E.~M.}\ \bibnamefont {Gonzalez}},
  \bibinfo {author} {\bibfnamefont {M.}~\bibnamefont {Velez}}, \bibinfo
  {author} {\bibfnamefont {J.~I.}\ \bibnamefont {Martin}}, \ and\ \bibinfo
  {author} {\bibfnamefont {J.~L.}\ \bibnamefont {Vicent}},\ }\href {\doibase
  10.1103/PhysRevB.82.174503} {\bibfield  {journal} {\bibinfo  {journal} {Phys.
  Rev. B}\ }\textbf {\bibinfo {volume} {82}},\ \bibinfo {pages} {174503}
  (\bibinfo {year} {2010})}\BibitemShut {NoStop}%
\bibitem [{\citenamefont {Perez~de Lara}\ \emph {et~al.}(2011)\citenamefont
  {Perez~de Lara}, \citenamefont {Erekhinsky}, \citenamefont {Gonzalez},
  \citenamefont {Rosen}, \citenamefont {Schuller},\ and\ \citenamefont
  {Vicent}}]{Lar11prb}%
  \BibitemOpen
  \bibfield  {author} {\bibinfo {author} {\bibfnamefont {D.}~\bibnamefont
  {Perez~de Lara}}, \bibinfo {author} {\bibfnamefont {M.}~\bibnamefont
  {Erekhinsky}}, \bibinfo {author} {\bibfnamefont {E.~M.}\ \bibnamefont
  {Gonzalez}}, \bibinfo {author} {\bibfnamefont {Y.~J.}\ \bibnamefont {Rosen}},
  \bibinfo {author} {\bibfnamefont {I.~K.}\ \bibnamefont {Schuller}}, \ and\
  \bibinfo {author} {\bibfnamefont {J.~L.}\ \bibnamefont {Vicent}},\ }\href
  {\doibase 10.1103/PhysRevB.83.174507} {\bibfield  {journal} {\bibinfo
  {journal} {Phys. Rev. B}\ }\textbf {\bibinfo {volume} {83}},\ \bibinfo
  {pages} {174507} (\bibinfo {year} {2011})}\BibitemShut {NoStop}%
\bibitem [{\citenamefont {Arzola}\ \emph {et~al.}(2011)\citenamefont {Arzola},
  \citenamefont {Volke-Sep\'ulveda},\ and\ \citenamefont {Mateos}}]{Arz11prl}%
  \BibitemOpen
  \bibfield  {author} {\bibinfo {author} {\bibfnamefont {A.~V.}\ \bibnamefont
  {Arzola}}, \bibinfo {author} {\bibfnamefont {K.}~\bibnamefont
  {Volke-Sep\'ulveda}}, \ and\ \bibinfo {author} {\bibfnamefont {J.~L.}\
  \bibnamefont {Mateos}},\ }\href {\doibase 10.1103/PhysRevLett.106.168104}
  {\bibfield  {journal} {\bibinfo  {journal} {Phys. Rev. Lett.}\ }\textbf
  {\bibinfo {volume} {106}},\ \bibinfo {pages} {168104} (\bibinfo {year}
  {2011})}\BibitemShut {NoStop}%
\bibitem [{\citenamefont {Olson~Reichhardt}\ and\ \citenamefont
  {Reichhardt}(2005)}]{Ols05pcs}%
  \BibitemOpen
  \bibfield  {author} {\bibinfo {author} {\bibfnamefont {C.}~\bibnamefont
  {Olson~Reichhardt}}\ and\ \bibinfo {author} {\bibfnamefont {C.}~\bibnamefont
  {Reichhardt}},\ }\href {\doibase 10.1016/j.physc.2005.07.017} {\bibfield
  {journal} {\bibinfo  {journal} {Physica C: Superconductivity}\ }\textbf
  {\bibinfo {volume} {432}},\ \bibinfo {pages} {125 } (\bibinfo {year}
  {2005})}\BibitemShut {NoStop}%
\bibitem [{\citenamefont {Morrison}\ and\ \citenamefont
  {Rose}(1970)}]{Mor70prl}%
  \BibitemOpen
  \bibfield  {author} {\bibinfo {author} {\bibfnamefont {D.~D.}\ \bibnamefont
  {Morrison}}\ and\ \bibinfo {author} {\bibfnamefont {R.~M.}\ \bibnamefont
  {Rose}},\ }\href {\doibase 10.1103/PhysRevLett.25.356} {\bibfield  {journal}
  {\bibinfo  {journal} {Phys. Rev. Lett.}\ }\textbf {\bibinfo {volume} {25}},\
  \bibinfo {pages} {356} (\bibinfo {year} {1970})}\BibitemShut {NoStop}%
\bibitem [{\citenamefont {Utke}\ \emph {et~al.}(2008)\citenamefont {Utke},
  \citenamefont {Hoffmann},\ and\ \citenamefont {Melngailis}}]{Utk08vst}%
  \BibitemOpen
  \bibfield  {author} {\bibinfo {author} {\bibfnamefont {I.}~\bibnamefont
  {Utke}}, \bibinfo {author} {\bibfnamefont {P.}~\bibnamefont {Hoffmann}}, \
  and\ \bibinfo {author} {\bibfnamefont {J.}~\bibnamefont {Melngailis}},\
  }\href@noop {} {\bibfield  {journal} {\bibinfo  {journal} {J. Vac. Sci.
  Technol. B}\ }\textbf {\bibinfo {volume} {26}},\ \bibinfo {pages} {1197}
  (\bibinfo {year} {2008})}\BibitemShut {NoStop}%
\bibitem [{\citenamefont {Huth}(2010)}]{Hut10inp}%
  \BibitemOpen
  \bibfield  {author} {\bibinfo {author} {\bibfnamefont {M.}~\bibnamefont
  {Huth}},\ }in\ \href@noop {} {\emph {\bibinfo {booktitle} {Proceedings of
  Beilstein Symposium}}}\ (\bibinfo {year} {2010})\ pp.\ \bibinfo {pages}
  {193--211}\BibitemShut {NoStop}%
\bibitem [{\citenamefont {Dobrovolskiy}\ \emph
  {et~al.}(2012{\natexlab{a}})\citenamefont {Dobrovolskiy}, \citenamefont
  {Begun}, \citenamefont {Huth},\ and\ \citenamefont {Shklovskij}}]{Dob12njp}%
  \BibitemOpen
  \bibfield  {author} {\bibinfo {author} {\bibfnamefont {O.~V.}\ \bibnamefont
  {Dobrovolskiy}}, \bibinfo {author} {\bibfnamefont {E.}~\bibnamefont {Begun}},
  \bibinfo {author} {\bibfnamefont {M.}~\bibnamefont {Huth}}, \ and\ \bibinfo
  {author} {\bibfnamefont {V.~A.}\ \bibnamefont {Shklovskij}},\ }\href
  {http://stacks.iop.org/1367-2630/14/i=11/a=113027} {\bibfield  {journal}
  {\bibinfo  {journal} {New J. Phys.}\ }\textbf {\bibinfo {volume} {14}},\
  \bibinfo {pages} {113027} (\bibinfo {year} {2012}{\natexlab{a}})}\BibitemShut
  {NoStop}%
\bibitem [{\citenamefont {Huth}\ \emph {et~al.}(2012)\citenamefont {Huth},
  \citenamefont {Porrati}, \citenamefont {Schwalb}, \citenamefont {Winhold},
  \citenamefont {Sachser}, \citenamefont {Dukic}, \citenamefont {Adams},\ and\
  \citenamefont {Fantner}}]{Hut12bjn}%
  \BibitemOpen
  \bibfield  {author} {\bibinfo {author} {\bibfnamefont {M.}~\bibnamefont
  {Huth}}, \bibinfo {author} {\bibfnamefont {F.}~\bibnamefont {Porrati}},
  \bibinfo {author} {\bibfnamefont {C.}~\bibnamefont {Schwalb}}, \bibinfo
  {author} {\bibfnamefont {M.}~\bibnamefont {Winhold}}, \bibinfo {author}
  {\bibfnamefont {R.}~\bibnamefont {Sachser}}, \bibinfo {author} {\bibfnamefont
  {M.}~\bibnamefont {Dukic}}, \bibinfo {author} {\bibfnamefont
  {J.}~\bibnamefont {Adams}}, \ and\ \bibinfo {author} {\bibfnamefont
  {G.}~\bibnamefont {Fantner}},\ }\href {\doibase 10.3762/bjnano.3.70}
  {\bibfield  {journal} {\bibinfo  {journal} {Beilstein J. Nanotechnol.}\
  }\textbf {\bibinfo {volume} {3}},\ \bibinfo {pages} {597} (\bibinfo {year}
  {2012})}\BibitemShut {NoStop}%
\bibitem [{\citenamefont {Dobrovolskiy}\ \emph {et~al.}(2010)\citenamefont
  {Dobrovolskiy}, \citenamefont {Huth},\ and\ \citenamefont
  {Shklovskij}}]{Dob10sst}%
  \BibitemOpen
  \bibfield  {author} {\bibinfo {author} {\bibfnamefont {O.~V.}\ \bibnamefont
  {Dobrovolskiy}}, \bibinfo {author} {\bibfnamefont {M.}~\bibnamefont {Huth}},
  \ and\ \bibinfo {author} {\bibfnamefont {V.~A.}\ \bibnamefont {Shklovskij}},\
  }\href {\doibase doi:10.1088/0953-2048/23/12/125014} {\bibfield  {journal}
  {\bibinfo  {journal} {Supercond. Sci. Technol.}\ }\textbf {\bibinfo {volume}
  {23}},\ \bibinfo {pages} {125014} (\bibinfo {year} {2010})}\BibitemShut
  {NoStop}%
\bibitem [{\citenamefont {Dobrovolskiy}\ \emph {et~al.}(2011)\citenamefont
  {Dobrovolskiy}, \citenamefont {Begun}, \citenamefont {Huth}, \citenamefont
  {Shklovskij},\ and\ \citenamefont {Tsindlekht}}]{Dob11pcs}%
  \BibitemOpen
  \bibfield  {author} {\bibinfo {author} {\bibfnamefont {O.~V.}\ \bibnamefont
  {Dobrovolskiy}}, \bibinfo {author} {\bibfnamefont {E.}~\bibnamefont {Begun}},
  \bibinfo {author} {\bibfnamefont {M.}~\bibnamefont {Huth}}, \bibinfo {author}
  {\bibfnamefont {V.~A.}\ \bibnamefont {Shklovskij}}, \ and\ \bibinfo {author}
  {\bibfnamefont {M.~I.}\ \bibnamefont {Tsindlekht}},\ }\href {\doibase
  10.1016/j.physc.2011.05.245} {\bibfield  {journal} {\bibinfo  {journal}
  {Physica C}\ }\textbf {\bibinfo {volume} {471}},\ \bibinfo {pages} {449}
  (\bibinfo {year} {2011})}\BibitemShut {NoStop}%
\bibitem [{\citenamefont {Dobrovolskiy}\ \emph
  {et~al.}(2012{\natexlab{b}})\citenamefont {Dobrovolskiy}, \citenamefont
  {Huth},\ and\ \citenamefont {Shklovskij}}]{Dob12ppa}%
  \BibitemOpen
  \bibfield  {author} {\bibinfo {author} {\bibfnamefont {O.~V.}\ \bibnamefont
  {Dobrovolskiy}}, \bibinfo {author} {\bibfnamefont {M.}~\bibnamefont {Huth}},
  \ and\ \bibinfo {author} {\bibfnamefont {V.~A.}\ \bibnamefont {Shklovskij}},\
  }\href@noop {} {\bibfield  {journal} {\bibinfo  {journal} {Acta Phys. Pol.
  A}\ }\textbf {\bibinfo {volume} {121}},\ \bibinfo {pages} {82} (\bibinfo
  {year} {2012}{\natexlab{b}})}\BibitemShut {NoStop}%
\bibitem [{\citenamefont {Jung}\ \emph {et~al.}(1996)\citenamefont {Jung},
  \citenamefont {Kissner},\ and\ \citenamefont {H\"anggi}}]{Jun96prl}%
  \BibitemOpen
  \bibfield  {author} {\bibinfo {author} {\bibfnamefont {P.}~\bibnamefont
  {Jung}}, \bibinfo {author} {\bibfnamefont {J.~G.}\ \bibnamefont {Kissner}}, \
  and\ \bibinfo {author} {\bibfnamefont {P.}~\bibnamefont {H\"anggi}},\ }\href
  {\doibase 10.1103/PhysRevLett.76.3436} {\bibfield  {journal} {\bibinfo
  {journal} {Phys. Rev. Lett.}\ }\textbf {\bibinfo {volume} {76}},\ \bibinfo
  {pages} {3436} (\bibinfo {year} {1996})}\BibitemShut {NoStop}%
\bibitem [{\citenamefont {Mateos}(2000)}]{Mat00prl}%
  \BibitemOpen
  \bibfield  {author} {\bibinfo {author} {\bibfnamefont {J.~L.}\ \bibnamefont
  {Mateos}},\ }\href {\doibase 10.1103/PhysRevLett.84.258} {\bibfield
  {journal} {\bibinfo  {journal} {Phys. Rev. Lett.}\ }\textbf {\bibinfo
  {volume} {84}},\ \bibinfo {pages} {258} (\bibinfo {year} {2000})}\BibitemShut
  {NoStop}%
\bibitem [{\citenamefont {Reimann}\ and\ \citenamefont
  {H\"anggi}(2002)}]{Rei02apa}%
  \BibitemOpen
  \bibfield  {author} {\bibinfo {author} {\bibfnamefont {P.}~\bibnamefont
  {Reimann}}\ and\ \bibinfo {author} {\bibfnamefont {P.}~\bibnamefont
  {H\"anggi}},\ }\href {http://dx.doi.org/10.1007/s003390201331} {\bibfield
  {journal} {\bibinfo  {journal} {Appl. Phys. A}\ }\textbf {\bibinfo {volume}
  {75}},\ \bibinfo {pages} {169} (\bibinfo {year} {2002})}\BibitemShut
  {NoStop}%
\bibitem [{\citenamefont {Shklovskij}\ and\ \citenamefont
  {Sosedkin}(2009)}]{Shk09prb}%
  \BibitemOpen
  \bibfield  {author} {\bibinfo {author} {\bibfnamefont {V.~A.}\ \bibnamefont
  {Shklovskij}}\ and\ \bibinfo {author} {\bibfnamefont {V.~V.}\ \bibnamefont
  {Sosedkin}},\ }\href {\doibase 10.1103/PhysRevB.80.214526} {\bibfield
  {journal} {\bibinfo  {journal} {Phys. Rev. B}\ }\textbf {\bibinfo {volume}
  {80}},\ \bibinfo {pages} {214526} (\bibinfo {year} {2009})}\BibitemShut
  {NoStop}%
\bibitem [{\citenamefont {Shklovskij}\ and\ \citenamefont
  {Dobrovolskiy}(2011)}]{Shk11prb}%
  \BibitemOpen
  \bibfield  {author} {\bibinfo {author} {\bibfnamefont {V.~A.}\ \bibnamefont
  {Shklovskij}}\ and\ \bibinfo {author} {\bibfnamefont {O.~V.}\ \bibnamefont
  {Dobrovolskiy}},\ }\href {\doibase 10.1103/PhysRevB.84.054515} {\bibfield
  {journal} {\bibinfo  {journal} {Phys. Rev. B}\ }\textbf {\bibinfo {volume}
  {84}},\ \bibinfo {pages} {054515} (\bibinfo {year} {2011})}\BibitemShut
  {NoStop}%
\bibitem [{\citenamefont {Shklovskij}\ and\ \citenamefont
  {Dobrovolskiy}(2008)}]{Shk08prb}%
  \BibitemOpen
  \bibfield  {author} {\bibinfo {author} {\bibfnamefont {V.~A.}\ \bibnamefont
  {Shklovskij}}\ and\ \bibinfo {author} {\bibfnamefont {O.~V.}\ \bibnamefont
  {Dobrovolskiy}},\ }\href {\doibase 10.1103/PhysRevB.78.104526} {\bibfield
  {journal} {\bibinfo  {journal} {Phys. Rev. B}\ }\textbf {\bibinfo {volume}
  {78}},\ \bibinfo {pages} {104526} (\bibinfo {year} {2008})}\BibitemShut
  {NoStop}%
\bibitem [{\citenamefont {Martinoli}\ \emph {et~al.}(1990)\citenamefont
  {Martinoli}, \citenamefont {Fluckiger}, \citenamefont {Marsico},
  \citenamefont {Srivastava}, \citenamefont {Leemann},\ and\ \citenamefont
  {Gavilano}}]{Mar90pbc}%
  \BibitemOpen
  \bibfield  {author} {\bibinfo {author} {\bibfnamefont {P.}~\bibnamefont
  {Martinoli}}, \bibinfo {author} {\bibfnamefont {P.}~\bibnamefont
  {Fluckiger}}, \bibinfo {author} {\bibfnamefont {V.}~\bibnamefont {Marsico}},
  \bibinfo {author} {\bibfnamefont {P.~K.}\ \bibnamefont {Srivastava}},
  \bibinfo {author} {\bibfnamefont {C.}~\bibnamefont {Leemann}}, \ and\
  \bibinfo {author} {\bibfnamefont {J.~L.}\ \bibnamefont {Gavilano}},\ }\href
  {\doibase 10.1016/S0921-4526(09)80167-1} {\bibfield  {journal} {\bibinfo
  {journal} {Physica B}\ }\textbf {\bibinfo {volume} {165--166}},\ \bibinfo
  {pages} {1163} (\bibinfo {year} {1990})}\BibitemShut {NoStop}%
\bibitem [{\citenamefont {Coffey}\ and\ \citenamefont {Clem}(1991)}]{Cof91prl}%
  \BibitemOpen
  \bibfield  {author} {\bibinfo {author} {\bibfnamefont {M.~W.}\ \bibnamefont
  {Coffey}}\ and\ \bibinfo {author} {\bibfnamefont {J.~R.}\ \bibnamefont
  {Clem}},\ }\href {\doibase 10.1103/PhysRevLett.67.386} {\bibfield  {journal}
  {\bibinfo  {journal} {Phys. Rev. Lett.}\ }\textbf {\bibinfo {volume} {67}},\
  \bibinfo {pages} {386} (\bibinfo {year} {1991})}\BibitemShut {NoStop}%
\bibitem [{\citenamefont {Mawatari}(1999)}]{Maw99prb}%
  \BibitemOpen
  \bibfield  {author} {\bibinfo {author} {\bibfnamefont {Y.}~\bibnamefont
  {Mawatari}},\ }\href {\doibase 10.1103/PhysRevB.59.12033} {\bibfield
  {journal} {\bibinfo  {journal} {Phys. Rev. B}\ }\textbf {\bibinfo {volume}
  {59}},\ \bibinfo {pages} {12033} (\bibinfo {year} {1999})}\BibitemShut
  {NoStop}%
\bibitem [{\citenamefont {Golosovsky}\ \emph {et~al.}(1992)\citenamefont
  {Golosovsky}, \citenamefont {Naveh},\ and\ \citenamefont
  {Davidov}}]{Gol92prb}%
  \BibitemOpen
  \bibfield  {author} {\bibinfo {author} {\bibfnamefont {M.}~\bibnamefont
  {Golosovsky}}, \bibinfo {author} {\bibfnamefont {Y.}~\bibnamefont {Naveh}}, \
  and\ \bibinfo {author} {\bibfnamefont {D.}~\bibnamefont {Davidov}},\ }\href
  {\doibase 10.1103/PhysRevB.45.7495} {\bibfield  {journal} {\bibinfo
  {journal} {Phys. Rev. B}\ }\textbf {\bibinfo {volume} {45}},\ \bibinfo
  {pages} {7495} (\bibinfo {year} {1992})}\BibitemShut {NoStop}%
\bibitem [{\citenamefont {Bartussek}\ \emph {et~al.}(1994)\citenamefont
  {Bartussek}, \citenamefont {H\"anggi},\ and\ \citenamefont
  {Kissner}}]{Bar94epl}%
  \BibitemOpen
  \bibfield  {author} {\bibinfo {author} {\bibfnamefont {R.}~\bibnamefont
  {Bartussek}}, \bibinfo {author} {\bibfnamefont {P.}~\bibnamefont {H\"anggi}},
  \ and\ \bibinfo {author} {\bibfnamefont {J.~G.}\ \bibnamefont {Kissner}},\
  }\href {http://stacks.iop.org/0295-5075/28/i=7/a=001} {\bibfield  {journal}
  {\bibinfo  {journal} {Europhys. Lett.}\ }\textbf {\bibinfo {volume} {28}},\
  \bibinfo {pages} {459} (\bibinfo {year} {1994})}\BibitemShut {NoStop}%
\bibitem [{\citenamefont {H\"anggi}\ and\ \citenamefont
  {Bartussek}(1996)}]{Han96inc}%
  \BibitemOpen
  \bibfield  {author} {\bibinfo {author} {\bibfnamefont {P.}~\bibnamefont
  {H\"anggi}}\ and\ \bibinfo {author} {\bibfnamefont {R.}~\bibnamefont
  {Bartussek}},\ }in\ \href {http://dx.doi.org/10.1007/BFb0105447} {\emph
  {\bibinfo {booktitle} {Nonlinear Physics of Complex Systems}}},\ \bibinfo
  {series} {Lecture Notes in Physics}, Vol.\ \bibinfo {volume} {476},\ \bibinfo
  {editor} {edited by\ \bibinfo {editor} {\bibfnamefont {J.}~\bibnamefont
  {Parisi}}, \bibinfo {editor} {\bibfnamefont {S.}~\bibnamefont {M\"uller}}, \
  and\ \bibinfo {editor} {\bibfnamefont {W.}~\bibnamefont {Zimmermann}}}\
  (\bibinfo  {publisher} {Springer, Berlin Heidelberg},\ \bibinfo {year}
  {1996})\ pp.\ \bibinfo {pages} {294--308}\BibitemShut {NoStop}%
\bibitem [{\citenamefont {Popescu}\ \emph {et~al.}(2000)\citenamefont
  {Popescu}, \citenamefont {Arizmendi}, \citenamefont {Salas-Brito},\ and\
  \citenamefont {Family}}]{Pop00prl}%
  \BibitemOpen
  \bibfield  {author} {\bibinfo {author} {\bibfnamefont {M.~N.}\ \bibnamefont
  {Popescu}}, \bibinfo {author} {\bibfnamefont {C.~M.}\ \bibnamefont
  {Arizmendi}}, \bibinfo {author} {\bibfnamefont {A.~L.}\ \bibnamefont
  {Salas-Brito}}, \ and\ \bibinfo {author} {\bibfnamefont {F.}~\bibnamefont
  {Family}},\ }\href {\doibase 10.1103/PhysRevLett.85.3321} {\bibfield
  {journal} {\bibinfo  {journal} {Phys. Rev. Lett.}\ }\textbf {\bibinfo
  {volume} {85}},\ \bibinfo {pages} {3321} (\bibinfo {year}
  {2000})}\BibitemShut {NoStop}%
\bibitem [{\citenamefont {Coffey}\ \emph {et~al.}(2004)\citenamefont {Coffey},
  \citenamefont {Kalmykov},\ and\ \citenamefont {Waldron}}]{Cof04boo}%
  \BibitemOpen
  \bibfield  {author} {\bibinfo {author} {\bibfnamefont {W.~T.}\ \bibnamefont
  {Coffey}}, \bibinfo {author} {\bibfnamefont {Y.~P.}\ \bibnamefont
  {Kalmykov}}, \ and\ \bibinfo {author} {\bibfnamefont {J.~T.}\ \bibnamefont
  {Waldron}},\ }\href@noop {} {\emph {\bibinfo {title} {The Langevin
  Equation}}}\ (\bibinfo  {publisher} {World Scientific, Singapore},\ \bibinfo
  {year} {2004})\BibitemShut {NoStop}%
\bibitem [{\citenamefont {Soroka}\ \emph {et~al.}(2007)\citenamefont {Soroka},
  \citenamefont {Shklovskij},\ and\ \citenamefont {Huth}}]{Sor07prb}%
  \BibitemOpen
  \bibfield  {author} {\bibinfo {author} {\bibfnamefont {O.~K.}\ \bibnamefont
  {Soroka}}, \bibinfo {author} {\bibfnamefont {V.~A.}\ \bibnamefont
  {Shklovskij}}, \ and\ \bibinfo {author} {\bibfnamefont {M.}~\bibnamefont
  {Huth}},\ }\href {\doibase 10.1103/PhysRevB.76.014504} {\bibfield  {journal}
  {\bibinfo  {journal} {Phys. Rev. B}\ }\textbf {\bibinfo {volume} {76}},\
  \bibinfo {pages} {014504} (\bibinfo {year} {2007})}\BibitemShut {NoStop}%
\bibitem [{\citenamefont {Huth}\ \emph {et~al.}(2002)\citenamefont {Huth},
  \citenamefont {Ritley}, \citenamefont {Oster}, \citenamefont {Dosch},\ and\
  \citenamefont {Adrian}}]{Hut02afm}%
  \BibitemOpen
  \bibfield  {author} {\bibinfo {author} {\bibfnamefont {M.}~\bibnamefont
  {Huth}}, \bibinfo {author} {\bibfnamefont {K.}~\bibnamefont {Ritley}},
  \bibinfo {author} {\bibfnamefont {J.}~\bibnamefont {Oster}}, \bibinfo
  {author} {\bibfnamefont {H.}~\bibnamefont {Dosch}}, \ and\ \bibinfo {author}
  {\bibfnamefont {H.}~\bibnamefont {Adrian}},\ }\href {\doibase
  10.1002/1616-3028(20020517)12:5<333::AID-ADFM333>3.0.CO;2-C} {\bibfield
  {journal} {\bibinfo  {journal} {Adv. Func. Mat.}\ }\textbf {\bibinfo {volume}
  {12}},\ \bibinfo {pages} {333} (\bibinfo {year} {2002})}\BibitemShut
  {NoStop}%
\bibitem [{\citenamefont {Oster}\ \emph {et~al.}(2005)\citenamefont {Oster},
  \citenamefont {Kallmayer}, \citenamefont {Wiehl}, \citenamefont {Elmers},
  \citenamefont {Adrian}, \citenamefont {Porrati},\ and\ \citenamefont
  {Huth}}]{Ost05jap}%
  \BibitemOpen
  \bibfield  {author} {\bibinfo {author} {\bibfnamefont {J.}~\bibnamefont
  {Oster}}, \bibinfo {author} {\bibfnamefont {M.}~\bibnamefont {Kallmayer}},
  \bibinfo {author} {\bibfnamefont {L.}~\bibnamefont {Wiehl}}, \bibinfo
  {author} {\bibfnamefont {H.~J.}\ \bibnamefont {Elmers}}, \bibinfo {author}
  {\bibfnamefont {H.}~\bibnamefont {Adrian}}, \bibinfo {author} {\bibfnamefont
  {F.}~\bibnamefont {Porrati}}, \ and\ \bibinfo {author} {\bibfnamefont
  {M.}~\bibnamefont {Huth}},\ }\href {\doibase 10.1063/1.1825629} {\bibfield
  {journal} {\bibinfo  {journal} {J. Appl. Phys.}\ }\textbf {\bibinfo {volume}
  {97}},\ \bibinfo {eid} {014303} (\bibinfo {year} {2005})}\BibitemShut
  {NoStop}%
\bibitem [{\citenamefont {Shklovskij}\ and\ \citenamefont
  {Dobrovolskiy}(2012)}]{Shk12inb}%
  \BibitemOpen
  \bibfield  {author} {\bibinfo {author} {\bibfnamefont {V.~A.}\ \bibnamefont
  {Shklovskij}}\ and\ \bibinfo {author} {\bibfnamefont {O.~V.}\ \bibnamefont
  {Dobrovolskiy}},\ }\enquote {\bibinfo {title} {Microwave absorption by
  vortices in superconductors with a washboard pinning potential},}\ in\
  \href@noop {} {\emph {\bibinfo {booktitle} {Superconductors -- Materials,
  Properties and Applications}}},\ \bibinfo {editor} {edited by\ \bibinfo
  {editor} {\bibfnamefont {A.}~\bibnamefont {Gabovich}}}\ (\bibinfo
  {publisher} {InTech, Rijeka},\ \bibinfo {year} {2012})\ Chap.~\bibinfo
  {chapter} {11}, pp.\ \bibinfo {pages} {263--288}\BibitemShut {NoStop}%
\bibitem [{\citenamefont {Pompeo}\ \emph {et~al.}(2010)\citenamefont {Pompeo},
  \citenamefont {Silva}, \citenamefont {Sarti}, \citenamefont {Attanasio},\
  and\ \citenamefont {Cirillo}}]{Pom10pcs}%
  \BibitemOpen
  \bibfield  {author} {\bibinfo {author} {\bibfnamefont {N.}~\bibnamefont
  {Pompeo}}, \bibinfo {author} {\bibfnamefont {E.}~\bibnamefont {Silva}},
  \bibinfo {author} {\bibfnamefont {S.}~\bibnamefont {Sarti}}, \bibinfo
  {author} {\bibfnamefont {C.}~\bibnamefont {Attanasio}}, \ and\ \bibinfo
  {author} {\bibfnamefont {C.}~\bibnamefont {Cirillo}},\ }\href {\doibase
  10.1016/j.physc.2010.02.063} {\bibfield  {journal} {\bibinfo  {journal}
  {Physica C}\ }\textbf {\bibinfo {volume} {470}},\ \bibinfo {pages} {901}
  (\bibinfo {year} {2010})}\BibitemShut {NoStop}%
\bibitem [{\citenamefont {Janju\ifmmode \check{s}\else
  \v{s}\fi{}evi\ifmmode~\acute{c}\else \'{c}\fi{}}\ \emph
  {et~al.}(2006)\citenamefont {Janju\ifmmode \check{s}\else
  \v{s}\fi{}evi\ifmmode~\acute{c}\else \'{c}\fi{}}, \citenamefont
  {Grbi\ifmmode~\acute{c}\else \'{c}\fi{}}, \citenamefont
  {Po\ifmmode~\check{z}\else \v{z}\fi{}ek}, \citenamefont {Dul\ifmmode
  \check{c}\else \v{c}\fi{}i\ifmmode~\acute{c}\else \'{c}\fi{}}, \citenamefont
  {Paar}, \citenamefont {Nebendahl},\ and\ \citenamefont {Wagner}}]{Jan06prb}%
  \BibitemOpen
  \bibfield  {author} {\bibinfo {author} {\bibfnamefont {D.}~\bibnamefont
  {Janju\ifmmode \check{s}\else \v{s}\fi{}evi\ifmmode~\acute{c}\else
  \'{c}\fi{}}}, \bibinfo {author} {\bibfnamefont {M.~S.}\ \bibnamefont
  {Grbi\ifmmode~\acute{c}\else \'{c}\fi{}}}, \bibinfo {author} {\bibfnamefont
  {M.}~\bibnamefont {Po\ifmmode~\check{z}\else \v{z}\fi{}ek}}, \bibinfo
  {author} {\bibfnamefont {A.}~\bibnamefont {Dul\ifmmode \check{c}\else
  \v{c}\fi{}i\ifmmode~\acute{c}\else \'{c}\fi{}}}, \bibinfo {author}
  {\bibfnamefont {D.}~\bibnamefont {Paar}}, \bibinfo {author} {\bibfnamefont
  {B.}~\bibnamefont {Nebendahl}}, \ and\ \bibinfo {author} {\bibfnamefont
  {T.}~\bibnamefont {Wagner}},\ }\href@noop {} {\bibfield  {journal} {\bibinfo
  {journal} {Phys. Rev. B}\ }\textbf {\bibinfo {volume} {74}},\ \bibinfo
  {pages} {104501} (\bibinfo {year} {2006})}\BibitemShut {NoStop}%
\bibitem [{\citenamefont {Silva}\ \emph {et~al.}(2006)\citenamefont {Silva},
  \citenamefont {Pompeo}, \citenamefont {Sarti},\ and\ \citenamefont
  {Amabile}}]{Sil06inb}%
  \BibitemOpen
  \bibfield  {author} {\bibinfo {author} {\bibfnamefont {E.}~\bibnamefont
  {Silva}}, \bibinfo {author} {\bibfnamefont {N.}~\bibnamefont {Pompeo}},
  \bibinfo {author} {\bibfnamefont {S.}~\bibnamefont {Sarti}}, \ and\ \bibinfo
  {author} {\bibfnamefont {C.}~\bibnamefont {Amabile}},\ }\enquote {\bibinfo
  {title} {Vortex state microwave response in superconducting cuprates},}\ in\
  \href {http://arxiv.org/abs/cond-mat/0607676} {\emph {\bibinfo {booktitle}
  {Recent Developments in Superconductivity Research}}}\ (\bibinfo  {publisher}
  {Nova Science, Hauppauge, NY},\ \bibinfo {year} {2006})\ Chap.~\bibinfo
  {chapter} {1}, pp.\ \bibinfo {pages} {201--243}\BibitemShut {NoStop}%
\bibitem [{\citenamefont {Likharev}(1986)}]{Lik86boo}%
  \BibitemOpen
  \bibfield  {author} {\bibinfo {author} {\bibfnamefont {K.~K.}\ \bibnamefont
  {Likharev}},\ }\href@noop {} {\emph {\bibinfo {title} {Dynamics of Josephson
  Junctions and Circuits}}}\ (\bibinfo  {publisher} {Gordon and Breach, New
  York},\ \bibinfo {year} {1986})\BibitemShut {NoStop}%
\bibitem [{\citenamefont {W\"ordenweber}\ \emph {et~al.}(2012)\citenamefont
  {W\"ordenweber}, \citenamefont {Hollmann}, \citenamefont {Schubert},
  \citenamefont {Kutzner},\ and\ \citenamefont {Panaitov}}]{Wor12prb}%
  \BibitemOpen
  \bibfield  {author} {\bibinfo {author} {\bibfnamefont {R.}~\bibnamefont
  {W\"ordenweber}}, \bibinfo {author} {\bibfnamefont {E.}~\bibnamefont
  {Hollmann}}, \bibinfo {author} {\bibfnamefont {J.}~\bibnamefont {Schubert}},
  \bibinfo {author} {\bibfnamefont {R.}~\bibnamefont {Kutzner}}, \ and\
  \bibinfo {author} {\bibfnamefont {G.}~\bibnamefont {Panaitov}},\ }\href
  {\doibase 10.1103/PhysRevB.85.064503} {\bibfield  {journal} {\bibinfo
  {journal} {Phys. Rev. B}\ }\textbf {\bibinfo {volume} {85}},\ \bibinfo
  {pages} {064503} (\bibinfo {year} {2012})}\BibitemShut {NoStop}%
\bibitem [{\citenamefont {Perez~de Lara}\ \emph {et~al.}(2009)\citenamefont
  {Perez~de Lara}, \citenamefont {Casta\~no}, \citenamefont {Ng}, \citenamefont
  {Korner}, \citenamefont {Dumas}, \citenamefont {Gonzalez}, \citenamefont
  {Liu}, \citenamefont {Ross}, \citenamefont {Schuller},\ and\ \citenamefont
  {Vicent}}]{Lar09prb}%
  \BibitemOpen
  \bibfield  {author} {\bibinfo {author} {\bibfnamefont {D.}~\bibnamefont
  {Perez~de Lara}}, \bibinfo {author} {\bibfnamefont {F.~J.}\ \bibnamefont
  {Casta\~no}}, \bibinfo {author} {\bibfnamefont {B.~G.}\ \bibnamefont {Ng}},
  \bibinfo {author} {\bibfnamefont {H.~S.}\ \bibnamefont {Korner}}, \bibinfo
  {author} {\bibfnamefont {R.~K.}\ \bibnamefont {Dumas}}, \bibinfo {author}
  {\bibfnamefont {E.~M.}\ \bibnamefont {Gonzalez}}, \bibinfo {author}
  {\bibfnamefont {K.}~\bibnamefont {Liu}}, \bibinfo {author} {\bibfnamefont
  {C.~A.}\ \bibnamefont {Ross}}, \bibinfo {author} {\bibfnamefont {I.~K.}\
  \bibnamefont {Schuller}}, \ and\ \bibinfo {author} {\bibfnamefont {J.~L.}\
  \bibnamefont {Vicent}},\ }\href {\doibase 10.1103/PhysRevB.80.224510}
  {\bibfield  {journal} {\bibinfo  {journal} {Phys. Rev. B}\ }\textbf {\bibinfo
  {volume} {80}},\ \bibinfo {pages} {224510} (\bibinfo {year}
  {2009})}\BibitemShut {NoStop}%
\bibitem [{\citenamefont {Shklovskij}(1980)}]{Shk80ltp}%
  \BibitemOpen
  \bibfield  {author} {\bibinfo {author} {\bibfnamefont {V.~A.}\ \bibnamefont
  {Shklovskij}},\ }\href@noop {} {\bibfield  {journal} {\bibinfo  {journal} {J.
  Low Temp. Phys.}\ }\textbf {\bibinfo {volume} {41}},\ \bibinfo {pages} {375}
  (\bibinfo {year} {1980})}\BibitemShut {NoStop}%
\bibitem [{\citenamefont {Bezuglyj}\ and\ \citenamefont
  {Shklovskij}(1992)}]{Bez92pcs}%
  \BibitemOpen
  \bibfield  {author} {\bibinfo {author} {\bibfnamefont {A.}~\bibnamefont
  {Bezuglyj}}\ and\ \bibinfo {author} {\bibfnamefont {V.}~\bibnamefont
  {Shklovskij}},\ }\href {\doibase 10.1016/0921-4534(92)90165-9} {\bibfield
  {journal} {\bibinfo  {journal} {Physica C}\ }\textbf {\bibinfo {volume}
  {202}},\ \bibinfo {pages} {234} (\bibinfo {year} {1992})}\BibitemShut
  {NoStop}%
\bibitem [{\citenamefont {Leo}\ \emph {et~al.}(2011)\citenamefont {Leo},
  \citenamefont {Grimaldi}, \citenamefont {Citro}, \citenamefont {Nigro},
  \citenamefont {Pace},\ and\ \citenamefont {Huebener}}]{Leo11prb}%
  \BibitemOpen
  \bibfield  {author} {\bibinfo {author} {\bibfnamefont {A.}~\bibnamefont
  {Leo}}, \bibinfo {author} {\bibfnamefont {G.}~\bibnamefont {Grimaldi}},
  \bibinfo {author} {\bibfnamefont {R.}~\bibnamefont {Citro}}, \bibinfo
  {author} {\bibfnamefont {A.}~\bibnamefont {Nigro}}, \bibinfo {author}
  {\bibfnamefont {S.}~\bibnamefont {Pace}}, \ and\ \bibinfo {author}
  {\bibfnamefont {R.~P.}\ \bibnamefont {Huebener}},\ }\href@noop {} {\bibfield
  {journal} {\bibinfo  {journal} {Phys. Rev. B}\ }\textbf {\bibinfo {volume}
  {84}},\ \bibinfo {pages} {014536} (\bibinfo {year} {2011})}\BibitemShut
  {NoStop}%
\bibitem [{\citenamefont {Liang}\ and\ \citenamefont
  {Kunchur}(2010)}]{Lik10prb}%
  \BibitemOpen
  \bibfield  {author} {\bibinfo {author} {\bibfnamefont {M.}~\bibnamefont
  {Liang}}\ and\ \bibinfo {author} {\bibfnamefont {M.~N.}\ \bibnamefont
  {Kunchur}},\ }\href {\doibase 10.1103/PhysRevB.82.144517} {\bibfield
  {journal} {\bibinfo  {journal} {Phys. Rev. B}\ }\textbf {\bibinfo {volume}
  {82}},\ \bibinfo {pages} {144517} (\bibinfo {year} {2010})}\BibitemShut
  {NoStop}%
\bibitem [{\citenamefont {Gr\"uner}(1988)}]{Gru88rmp}%
  \BibitemOpen
  \bibfield  {author} {\bibinfo {author} {\bibfnamefont {G.}~\bibnamefont
  {Gr\"uner}},\ }\href {\doibase 10.1103/RevModPhys.60.1129} {\bibfield
  {journal} {\bibinfo  {journal} {Rev. Mod. Phys.}\ }\textbf {\bibinfo {volume}
  {60}},\ \bibinfo {pages} {1129} (\bibinfo {year} {1988})}\BibitemShut
  {NoStop}%
\bibitem [{\citenamefont {Golubov}\ \emph {et~al.}(2004)\citenamefont
  {Golubov}, \citenamefont {Kupriyanov},\ and\ \citenamefont
  {Il'ichev}}]{Gol04rmp}%
  \BibitemOpen
  \bibfield  {author} {\bibinfo {author} {\bibfnamefont {A.~A.}\ \bibnamefont
  {Golubov}}, \bibinfo {author} {\bibfnamefont {M.~Y.}\ \bibnamefont
  {Kupriyanov}}, \ and\ \bibinfo {author} {\bibfnamefont {E.}~\bibnamefont
  {Il'ichev}},\ }\href {\doibase 10.1103/RevModPhys.76.411} {\bibfield
  {journal} {\bibinfo  {journal} {Rev. Mod. Phys.}\ }\textbf {\bibinfo {volume}
  {76}},\ \bibinfo {pages} {411} (\bibinfo {year} {2004})}\BibitemShut
  {NoStop}%
\bibitem [{\citenamefont {Coffey}\ \emph {et~al.}(2000)\citenamefont {Coffey},
  \citenamefont {D\'ejardin},\ and\ \citenamefont {Kalmykov}}]{Cof00prb}%
  \BibitemOpen
  \bibfield  {author} {\bibinfo {author} {\bibfnamefont {W.~T.}\ \bibnamefont
  {Coffey}}, \bibinfo {author} {\bibfnamefont {J.~L.}\ \bibnamefont
  {D\'ejardin}}, \ and\ \bibinfo {author} {\bibfnamefont {Y.~P.}\ \bibnamefont
  {Kalmykov}},\ }\href {\doibase 10.1103/PhysRevB.62.3480} {\bibfield
  {journal} {\bibinfo  {journal} {Phys. Rev. B}\ }\textbf {\bibinfo {volume}
  {62}},\ \bibinfo {pages} {3480} (\bibinfo {year} {2000})}\BibitemShut
  {NoStop}%
\bibitem [{\citenamefont {Titov}\ \emph {et~al.}(2010)\citenamefont {Titov},
  \citenamefont {D\'ejardin}, \citenamefont {El~Mrabti},\ and\ \citenamefont
  {Kalmykov}}]{Tit10prb}%
  \BibitemOpen
  \bibfield  {author} {\bibinfo {author} {\bibfnamefont {S.~V.}\ \bibnamefont
  {Titov}}, \bibinfo {author} {\bibfnamefont {P.-M.}\ \bibnamefont
  {D\'ejardin}}, \bibinfo {author} {\bibfnamefont {H.}~\bibnamefont
  {El~Mrabti}}, \ and\ \bibinfo {author} {\bibfnamefont {Y.~P.}\ \bibnamefont
  {Kalmykov}},\ }\href {\doibase 10.1103/PhysRevB.82.100413} {\bibfield
  {journal} {\bibinfo  {journal} {Phys. Rev. B}\ }\textbf {\bibinfo {volume}
  {82}},\ \bibinfo {pages} {100413} (\bibinfo {year} {2010})}\BibitemShut
  {NoStop}%
\bibitem [{\citenamefont {Gammaitoni}\ \emph {et~al.}(1998)\citenamefont
  {Gammaitoni}, \citenamefont {H\"anggi}, \citenamefont {Jung},\ and\
  \citenamefont {Marchesoni}}]{Gam98rmp}%
  \BibitemOpen
  \bibfield  {author} {\bibinfo {author} {\bibfnamefont {L.}~\bibnamefont
  {Gammaitoni}}, \bibinfo {author} {\bibfnamefont {P.}~\bibnamefont
  {H\"anggi}}, \bibinfo {author} {\bibfnamefont {P.}~\bibnamefont {Jung}}, \
  and\ \bibinfo {author} {\bibfnamefont {F.}~\bibnamefont {Marchesoni}},\
  }\href {\doibase 10.1103/RevModPhys.70.223} {\bibfield  {journal} {\bibinfo
  {journal} {Rev. Mod. Phys.}\ }\textbf {\bibinfo {volume} {70}},\ \bibinfo
  {pages} {223} (\bibinfo {year} {1998})}\BibitemShut {NoStop}%
\bibitem [{\citenamefont {Bogu\~n\'a}\ \emph {et~al.}(1998)\citenamefont
  {Bogu\~n\'a}, \citenamefont {Porr\`a}, \citenamefont {Masoliver},\ and\
  \citenamefont {Lindenberg}}]{Bog98pre}%
  \BibitemOpen
  \bibfield  {author} {\bibinfo {author} {\bibfnamefont {M.}~\bibnamefont
  {Bogu\~n\'a}}, \bibinfo {author} {\bibfnamefont {J.~M.}\ \bibnamefont
  {Porr\`a}}, \bibinfo {author} {\bibfnamefont {J.}~\bibnamefont {Masoliver}},
  \ and\ \bibinfo {author} {\bibfnamefont {K.}~\bibnamefont {Lindenberg}},\
  }\href {\doibase 10.1103/PhysRevE.57.3990} {\bibfield  {journal} {\bibinfo
  {journal} {Phys. Rev. E}\ }\textbf {\bibinfo {volume} {57}},\ \bibinfo
  {pages} {3990} (\bibinfo {year} {1998})}\BibitemShut {NoStop}%
\bibitem [{\citenamefont {Mantegna}\ and\ \citenamefont
  {Spagnolo}(1996)}]{Man96prl}%
  \BibitemOpen
  \bibfield  {author} {\bibinfo {author} {\bibfnamefont {R.~N.}\ \bibnamefont
  {Mantegna}}\ and\ \bibinfo {author} {\bibfnamefont {B.}~\bibnamefont
  {Spagnolo}},\ }\href {\doibase 10.1103/PhysRevLett.76.563} {\bibfield
  {journal} {\bibinfo  {journal} {Phys. Rev. Lett.}\ }\textbf {\bibinfo
  {volume} {76}},\ \bibinfo {pages} {563} (\bibinfo {year} {1996})}\BibitemShut
  {NoStop}%
\end{thebibliography}
\end{document}